\documentclass{article}
\usepackage{arxiv}
\usepackage{amsmath,amssymb,amsfonts}
\usepackage{algorithmic}
\usepackage{graphicx}
\usepackage{hyperref}
\usepackage{url}
\usepackage{textcomp}
\newcommand\norm[1]{\left\lVert#1\right\rVert}
\usepackage{multirow}
\usepackage{subfig}
\usepackage{booktabs}
\usepackage{tabularx}
\usepackage{tikz}
\usepackage{xcolor}
\usepackage{siunitx}
\usepackage[square,numbers]{natbib}
\usepackage{color}
\usepackage{enumitem}
\usepackage{footnote}
\usepackage{stfloats}
\usepackage{courier}
\usepackage[T1]{fontenc}
\usepackage{float}
\floatstyle{plaintop}
\restylefloat{table}
\usepackage{makecell}
\usepackage{pifont}
\usepackage{array}
\newcolumntype{P}[1]{>{\centering\arraybackslash}p{#1}}
\newcolumntype{L}{>{$}l<{$}}
\newcolumntype{C}{>{$}c<{$}}
\newcolumntype{R}{>{$}r<{$}}
\newcommand{\ie}{i.e., }

\newcommand{\rom}[1]{\uppercase\expandafter{\romannumeral #1\relax}}
\newcommand{\name}{NUNet}
\newcommand{\baselineNN}{SURFNet}
\newcommand{\Rey}{\mbox{\textit{Re}}}
\def\BibTeX{{\rm B\kern-.05em{\sc i\kern-.025em b}\kern-.08em
    T\kern-.1667em\lower.7ex\hbox{E}\kern-.125emX}}

\author{ {\hspace{1mm}Octavi Obiols-Sales} \\
	University of California, Irvine\\
	Irvine, CA\\
	\texttt{oobiols@uci.edu} \\
	\And
	{\hspace{1mm}Abhinav Vishnu} \\
	Advanced Micro Devices, Inc.\\
	Austin, TX \\
	\texttt{abhinav.vishnu@amd.com} \\
	\And
	{\hspace{1mm}Nicholas Malaya} \\
	Advanced Micro Devices, Inc.\\
	Austin, TX \\
	\texttt{nicholas.malaya@amd.com} \\
	\And
	{\hspace{1mm}Aparna Chandramowlishwaran} \\
	University of California, Irvine\\
	Irvine, CA \\
	\texttt{amowli@uci.edu} \\
}
\begin{document}

\title{\name: Deep Learning for Non-Uniform Super-Resolution of Turbulent Flows}
\maketitle

\begin{abstract}
  Deep Learning (DL) algorithms are becoming increasingly popular for the
  reconstruction of high-resolution turbulent flows (\emph{aka}
  super-resolution).  However, current DL approaches perform spatially
  \emph{uniform} super-resolution -- a key performance limiter for
  scalability of DL-based surrogates for Computational Fluid Dynamics (CFD).

To address the above challenge, we introduce \name, a deep
	learning-based adaptive mesh refinement (AMR) framework for non-uniform
  super-resolution of turbulent flows.  \name\ divides the input low-resolution
  flow field into \emph{patches}, scores each patch, and predicts
  their target resolution.  As a result, it outputs a spatially non-uniform flow
  field, adaptively refining regions of the fluid domain to achieve the target
  accuracy. We train \name\ with Reynolds-Averaged Navier-Stokes (RANS)
  solutions from three different canonical flows, namely turbulent channel
  flow, flat plate, and flow around ellipses. \name\ shows remarkable
  discerning properties, refining areas with complex flow features, such as
  near-wall domains and the wake region in flow around solid bodies, while
  leaving areas with smooth variations (such as the freestream) in the
  low-precision range. Hence, \name\ demonstrates an excellent qualitative and quantitative
  alignment with the traditional OpenFOAM AMR solver. Moreover, it reaches the
  same convergence guarantees as the AMR solver while accelerating it by
  $3.2-5.5\times$, including unseen-during-training geometries and boundary
  conditions, demonstrating its generalization capacities.  Due to \name's
  ability to super-resolve only regions of interest, it predicts the same
  target $1024\times1024$ spatial resolution $7-28.5\times$ faster than
  state-of-the-art DL methods and reduces the memory usage by $4.4-7.65\times$,
  showcasing improved scalability. 
\end{abstract}

\section{Introduction}

Computational Fluid Dynamics (CFD) is the \emph{de-facto} method for solving
the Navier-Stokes equations, a set of Partial Differential Equations (PDEs),
both for laminar and turbulent flow
problems~\cite{pope,malaya,openfoam,RANS,behnam,Mostafazadeh:aa}. Practical
turbulent CFD simulations require high spatial resolutions (such as 1024
$\times$ 1024~\cite{propergrid}), making these simulations computationally
expensive. There are widespread efforts to address this challenge and improve
the performance and scalability of solving these
systems~\cite{multigrid,salvadore,trying} for faster design space exploration.  Inspired by the
remarkable success of deep learning (DL) algorithms in both computer vision
(CV)~\cite{ViT} and natural language processing
(NLP)~\cite{BERT}, recent works have leveraged DL algorithms for
accelerating CFD simulations via \emph{super-resolution}, that is,
reconstructing expensive high-resolution (HR) solutions from their
cost-effective low-resolution (LR)
counterpart~\cite{meshfreeflownet,denoisingsuperresolution,JFMsuperresolution2021,surfnet,JFMsuperresolution2019}.

\begin{figure}[htbp]
\centering
\small
\includegraphics[width=0.8\columnwidth]{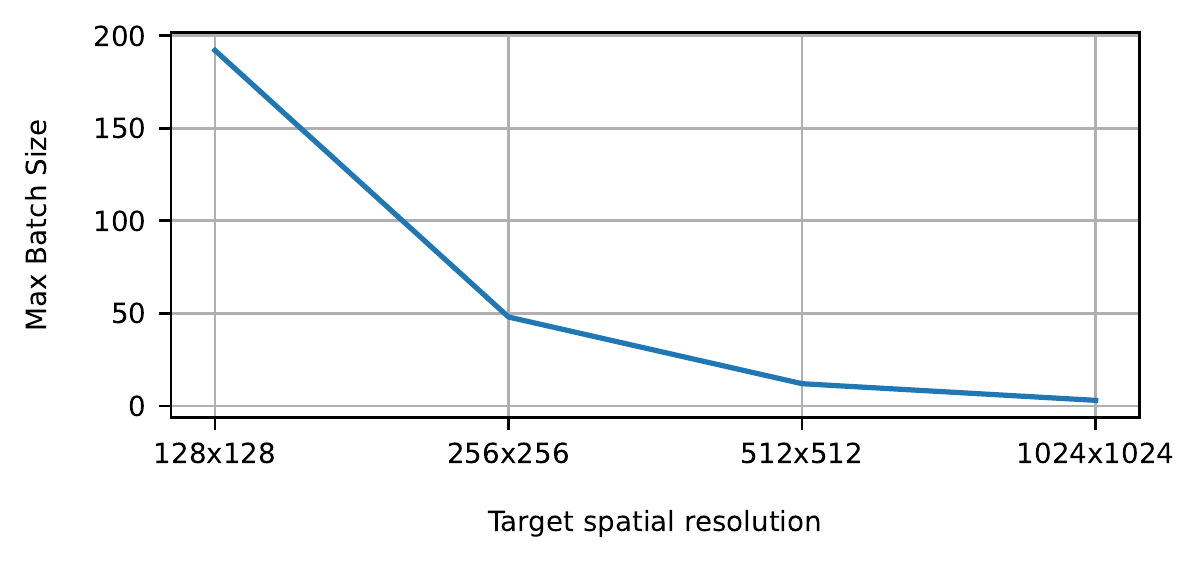}
	\caption{\small Maximum allowable batch size during inference
	at different target spatial resolutions for SOTA DL super-resolution
	methods~\cite{surfnet} on a 16GB NVIDIA V100 GPU with 16-bit floating point precision.\label{fig:intro}}
\end{figure}

State-of-the-art (SOTA) DL models for super-resolution have shown promise as
domain-agnostic models and real-time predictors that can generalize to a broad
set of flow configurations and conditions.  However, they all suffer from the
fundamental limitation of performing \emph{uniform} super-resolution,
that is, \emph{every} pixel of the input LR image is refined to the target
high resolution. As a result, SOTA~\cite{surfnet,meshfreeflownet,JFMsuperresolution2021}
methods for super-resolution need higher
computational resources - increased inference times 
and memory requirements - since the target high-resolution solution
is output in the entire physical domain.
Figure~\ref{fig:intro} shows the maximum
allowable batch size with increasing target spatial resolution of these methods. 
On a 16GB NVIDIA V100 GPU, these approaches do not allow more than
2 samples per batch during inference at high spatial resolutions,
such as $1024\times1024$, where more aspects of the physical phenomena
can be modeled. This severely limits the deployment of DL methods
for accelerating design space exploration in CFD. 

Spatially uniform outputs are computationally inefficient for two other reasons. 
First, they can under-resolve areas with complex flow features and
over-resolve regions with smooth fluctuations in the flow properties. It is
critical to capture this versatility for complex systems with large variations in their
local solutions, such as turbulent flows. Second,
current uniform super-resolution approaches need to know the target resolution \emph{a priori}.
As a result, they require a large number of HR labels at that specific resolution.
They hence need to rely either on publicly available datasets or on performing data collection.
Since the resolutions of the publicly available datasets are very limited, 
the majority of works~\cite{meshfreeflownet, JFMsuperresolution2021,
neuraloperator, liusuperresolution} end up performing computationally challenging high-resolution data collection.

Due to the large scale of many applications,
it is often infeasible to solve the problem on a uniform mesh to achieve the desired accuracy. 
For this reason, traditional numerical solvers do not refine the entire mesh but do so adaptively, 
refining only regions of strong flow variability for scalability and performance - 
a method commonly referred to as Adaptive Mesh Refinement (AMR)~\cite{ENZO,AMReX2021}.
However, traditional AMR methods in CFD suffer from two fundamental
limitations. First, a high degree of user intervention in the refining/coarsening decisions:
these decisions are based on heuristics that require problem-specific knowledge
and do not generalize well. Second, the mesh is refined iteratively, requiring
more compute time and memory compared to direct methods. 

\begin{figure}[htbp]
\centering
\small
\includegraphics[width=1\columnwidth]{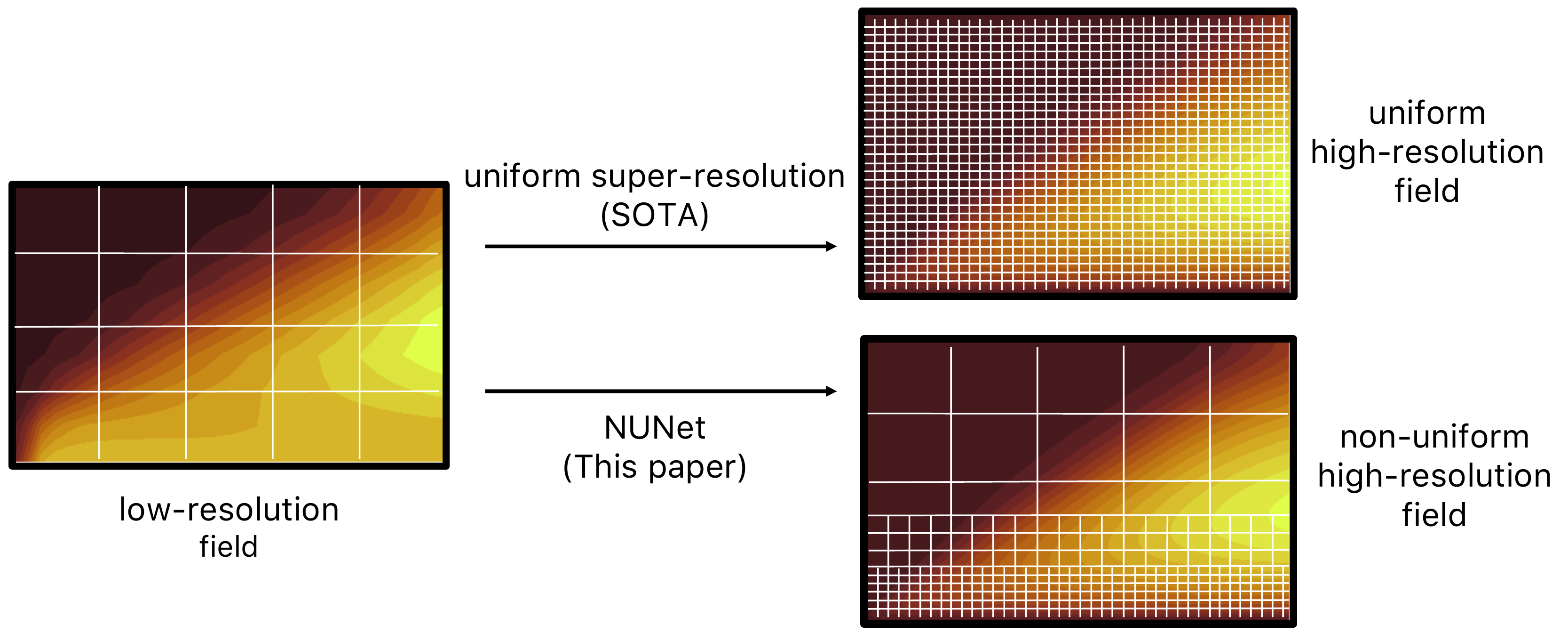}
	\caption{\small Current DL algorithms for super-resolution output the solution on a uniform fine mesh
	(top). Our objective with \name\ is
	to predict a spatially non-uniform output where only areas that require higher accuracy are refined (bottom). Hence, \name\ requires less compute time and memory resources while achieving the target accuracy.\label{fig:thispaper}}
\end{figure}

In this paper, we tackle all these challenges and present \name, a novel DL framework for \textbf{N}on-\textbf{U}niform
super-resolution. \name\ takes as input a LR flow field and outputs,
in one-shot, its final
non-uniform HR solution, as seen in Figure~\ref{fig:thispaper}.
Since only regions in areas that present complex
flow phenomena are refined, it requires less computational
resources. This enables larger batch sizes during inference 
at high spatial resolutions while reaching the target accuracy
compared to SOTA methods for super-resolution.
\name\ distinguishes between different regions of the domain by dividing the
input LR flow field into fixed-size \emph{patches}, adaptively increasing or
maintaining the spatial resolution of each patch, and predicting
a non-uniform HR flow field. We present \name\
as an end-to-end DL-physics solver framework where the non-uniform output field
from the model inference is input to a traditional physics solver that drives
the solution to convergence.  As a result, \name\ meets the same convergence
guarantees as AMR solvers which is critical for
practitioners~\cite{cfdnet,surfnet,maulik}.

Specifically, \name\ makes the following contributions:
\begin{itemize}

\item \textbf{Non-uniform super-resolution}. To enable non-uniform super-resolution, we propose a
novel \emph{scorer}-\emph{ranker}-\emph{decoder} DL algorithm, where the \emph{scorer} finds the spatial score of each patch,
the \emph{ranker} places each patch in its corresponding bin based on its score (which determines the target resolution of a patch), and
the \emph{decoder} reconstructs every patch in each bin to its final 
target resolution using semi-supervised learning.

\item \textbf{Requires minimal user intervention}. \name's training
	is semi-supervised - the loss function that guides
its optimization process is formed by 
		the governing equations of the problem - which poses
a two-fold advantage. First, the refinement or coarsening decisions are based 
on physics principles with the least possible human intervention,
as opposed to traditional AMR solvers that require
expert, domain, and even problem-specific knowledge
and a high degree of user intervention~\cite{AMReX2021}. 
From one single training, \name\ reaches SOTA refining/coarsening
decisions for different flow problems that exhibit very different flow phenomena,
showcasing remarkable generalization properties.
Second, \name\ does not require to know the target resolution a priori
and hence eliminates the need of expensive high-resolution data collection. 

\item \textbf{Outperforms traditional AMR solvers}. We evaluate \name\ 
on three canonical turbulent flows obtained from
Reynolds-Averaged Navier-Stokes (RANS) simulations on seven flow 
configurations unseen during training --
three unseen geometries and four unseen boundary conditions. \name's adaptively refined meshes
show excellent flow discerning properties and
agreement with the baseline OpenFOAM AMR solver in refining regions of interest
such as near-wall areas in wall-bounded problems (channel flow, flat plate), 
in flow around smooth solid bodies (flow around an airfoil), 
and the wake region in flow around thick solid bodies (flow around a cylinder) while maintaining less complex
flow areas, such as the freestream, at low resolution.
\name\ predicts the non-uniformly refined final flow field in a
single inference step as opposed to traditional
AMR solvers that require several iterations. 
As a result, \name\ reaches the same convergence guarantees
as the AMR solver while accelerating it by $3.2-5.5\times$.

\item \textbf{Requires less computational resources.} 
Due to \name's ability
to only refine specific areas of the flow and
avoid high-resolution inferences
in the entire the domain, it achieves a speedup of $7-28.5\times$ and reduces the memory
usage by $4.4-7.65\times$ at $1024\times 1024$ spatial resolutions
compared to SOTA DL methods that perform uniform super-resolution while
reaching SOTA accuracies.

\end{itemize}

\section{Related Work}
\label{sec:related}
\textbf{Adaptive Mesh Refinement (AMR)}. AMR is
a popular technique that makes it feasible to solve problems that are intractable on uniform grids
and it has been widely applied in traditional finite volume-based solvers. When PDEs
are solved numerically, they are often limited to a pre-determined computational grid or mesh.
However, different areas of the domain can
require different precisions where
non-uniform grids are better suited.
AMR algorithms adaptively and dynamically identify regions that require finer resolution (such as discontinuities, steep gradients, and shocks) and refine or coarsen the mesh to achieve the target accuracy. 
Therefore, AMR can scale to resolutions that would otherwise be infeasible on uniform meshes resulting in increased computational efficiency and storage savings. 
Moreover, the adaptive strategy offers more control over the grid resolution compared to the fixed resolution of the static grid approaches.
The most popular AMR techniques apply to traditional 
finite volume/finite element based 
numerical solvers. Even though recent approaches
have notably pushed the scalability boundaries of these systems~\cite{trying,AMReX2021,ENZO}
their core strategy results from the early work of ~\citeauthor{bergerAMRhyperbolic}~\cite{bergerAMRhyperbolic},
who introduced \emph{local adaptive mesh refinement}. This algorithm starts with a coarse mesh
from which certain cells are marked for refinement according
to either a user-supplied criterion or based
on the Richardson extrapolation~\cite{richardsonextrapolation}.
The principle of marking cells for refinement is widespread. Two main approaches exist for identifying cells for refinement. 
First, adjoint-based AMR~\cite{adjointbasedAMR}, which estimates
the discretization error in each cell and adapts the mesh for lowering 
these errors.
However, the optimal rationale for error estimation remains unknown~\cite{AMRbohn}. 
Second, feature-based AMR~\cite{featurebasedAMR}, where the user
supplies the variables (or features) to track and refines the
computational cells 
that meet a user-defined value of those variables. Feature-based AMR
is the most popular approach due to
less challenging implementations and accurate results in a wide
range of problems. However, feature-based
AMR approaches require both a high
degree of user intervention and expert, domain, and even 
specific knowledge of the problem at hand, 
and therefore have poor
generalization properties. Existing AMR methods are based on a handful
of heuristics whose long-term
or general optimality remains unknown. To overcome
this limitation, ~\citeauthor{AMRRL}~\cite{AMRRL} designed the AMR procedure
as a Markov Decision Process. However,
the training is done with ground truth data generated
from analytical solutions and can not be
extended to turbulent flows. 

There have been recent attempts to perform DL-based AMR. 
In~\cite{deepxde}, the authors increase the number of
solution points in those areas where the residual is highest. 
However, this mesh-free method imposes the same refinement
heuristics as traditional physical solvers and hence suffers
from the same limitations. In~\cite{amrnet}, the authors
develop AMRNet, a CNN-based model that performs \emph{multi-resolution},
where the network outputs a uniform flow field at different resolutions.
Since the output is uniform there is no discrimination between 
different areas of the flow. As opposed to the above approaches,
we design a DL algorithm for AMR,
where the output is non-uniform and we do not impose any 
heuristic during the optimization process of the network. 
Instead, the training
is guided by the governing equations of the problem that inform refining decisions (described in Section~\ref{sec:amrnet:loss}).

\textbf{Super-resolution}. DL algorithms have shown impressive results
for super-resolution. We find super-resolution techniques applied to both CV and CFD problems.
Two main research directions exist in CV:
single-image super-resolution (SISR) and reference-based super-resolution (RefSR). However, both SISR and RefSR
have a target resolution that is both known a priori and uniform~\cite{dong2014sisr,ledig2017gansr,RefSR}.
In~\cite{RefSR}, the authors present the \emph{texture transformer}, where
the query, key, and value of their attention module are formed by upsampled and downsampled
images of the input image together with a reference image from which textures are extracted.
In~\cite{differentiablepatchselection}, the authors provide
a differentiable module that selects the most salient patches
of the input image for image classification.
However, the unselected patches are
unused. In this paper, we are interested in super-resolution,
and therefore we keep all patches that cover the entire domain.

In CFD, we also find successful super-resolution attempts. 
Recent works use CNNs as finite-dimensional maps
~\cite{JFMsuperresolution2019,JFMsuperresolution2021,liusuperresolution}.
However, these approaches know the target resolution a priori, perform uniform SR,
and require large amounts of HR labels. To eliminate the need for large amounts of HR labels,
authors in~\cite{surfnet} developed
\emph{SURFNet}, a transfer learning-based uniform
SR framework. This work reduces the HR data requirement by $15\times$ 
while achieving resolution invariance. However, it's also limited to uniform SR.
Mesh-free, resolution-invariant methods~\cite{meshfreeflownet,deeponet,fourieroperator,neuraloperator,graphkernelnetwork} 
are a potential alternative to finite-dimensional maps because they can query the solution at any point in the domain and hence
are prone to perform non-uniform SR. In~\cite{meshfreeflownet},
the authors developed \emph{MeshFreeFlowNet}, 
an efficient framework for 
super-resolution of turbulent 
flows that demonstrates improved accuracy compared to baseline models~\cite{unet}. 
~\citeauthor{deeponet}
introduced a \emph{neural
operator}, which provides 
a set of network parameters 
compatible with different discretizations
and hence exhibits resolution-invariance --
achieving constant accuracy across discretizations.
However, this class of methods do not intrinsically discriminate
between different regions of the flow and hence end up
yielding uniform output resolutions. Moreover, 
they also suffer from the limitation of extensive HR data collection.

In this paper, we present a semi-supervised DL algorithm that adaptively refines the input mesh and outputs a non-uniform HR flow field, improving both inference times and memory requirements for scaling to large problem sizes. 
\name\ does not require knowledge of the target resoluton a priori, 
hence eliminates both the need for collecting extensive HR training data 
and the dependence on existing datasets
that are limited to specific resolutions.

\section{\name: DL for AMR}
\label{sec:amrnet}
Our objective is three-fold. First, to predict fine-grid
turbulent flows from their coarse-grid counterpart only in the regions of
interest. 
Second, to design a DL algorithm for AMR where these areas to refine are identified
with the least possible user intervention. 
Third, to output a solution that meets the same convergence guarantees as classical AMR solvers.

In this section, we present \name, a novel DL framework for adaptive super-resolution.  We first describe in detail the
neural network architecture and then,
present our semi-supervised learning approach that leverages a hybrid loss
function. Finally, we
outline the end-to-end framework, which reconstructs a non-uniform HR flow field
while reaching the same convergence as the state-of-the-art AMR solvers. 

\subsection{Neural Network Architecture}
\label{sec:amrnet:dnn}
\begin{figure*}[htbp]
\centering
\includegraphics[width=0.9\textwidth]{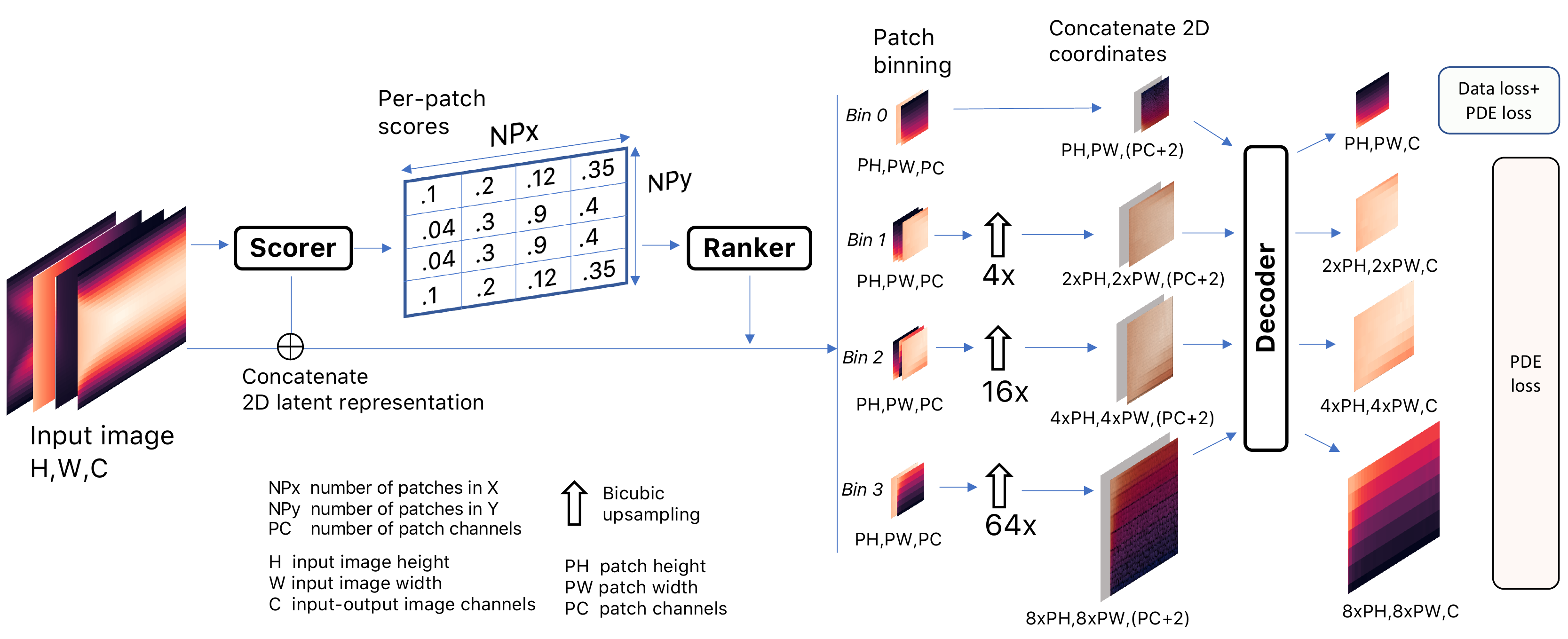}
	\caption{\small \name's DNN. The input is a four-channel
	LR image where each channel represents one flow variable.
	The LR image is first input to the \emph{scorer}, 
	which divides it into patches and outputs
	the score of each patch.
	The \emph{ranker} uses these scores to assign each patch to a bin, which is subsequently upsampled using a bicubic interpolation to its target resolution. Then, the upsampled patches are concatenated with their coordinates. 
	Finally, the \emph{decoder} maps this upsampled, intermediate representation to the final values of each patch. \name's DNN's output is multiple,
        consisting of a list of four-channel images at different spatial resolutions.
	\label{fig:amrnetdnn}}
\end{figure*}

We choose a deep neural network (DNN)
for the task of non-uniform super-resolution. 
The input to the DNN is a low-resolution
flow field, which is divided into fixed-size regions or \emph{patches}.
The output is a non-uniform resolution
flow field, where high-resolution is given only at specific \emph{patches} of the domain. 
The RANS equations with the Spalart-Allmaras model (described in Section~\ref{sec:exp:background}) predict four main flow variables -- mean x-velocity ($U$), mean y-velocity ($V$), 
the mean kinematic pressure ($p$), and the eddy viscosity ($\tilde{\nu}$).
Therefore, the input LR flow field consists of a four-channel tensor image where each channel
represents the values of one flow variable in the entire computational grid.
The DNN scores,
ranks (or bins), and predicts the target resolution of each patch. 
Figure~\ref{fig:amrnetdnn} illustrates the architecture composed of a \emph{scorer} network,
a \emph{ranker}, and a \emph{decoder} network.

\begin{figure}[htbp]
\centering
\small
\includegraphics[width=1\columnwidth]{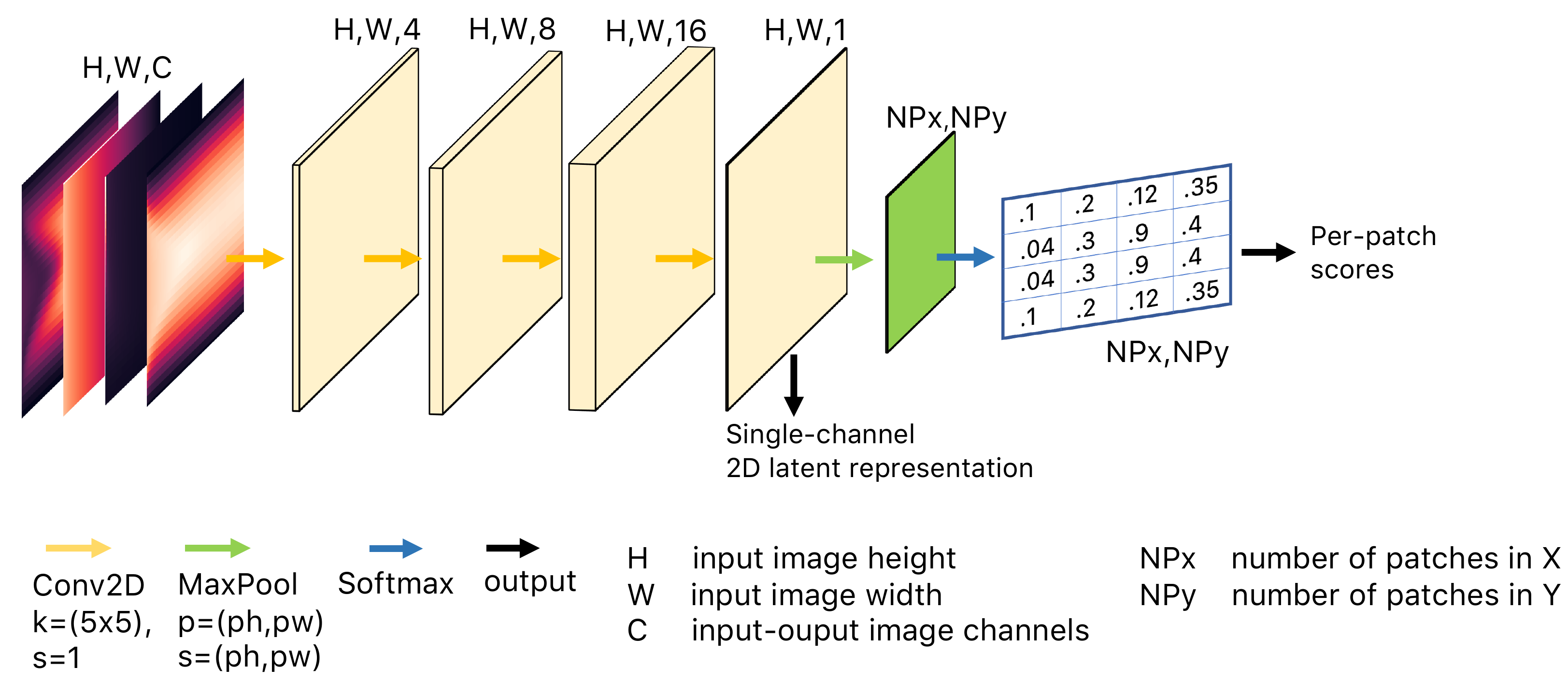}
	\caption{\small The \emph{scorer} network. It consists of four convolutional layers followed by maxpool and softmax layers. The first three convolutional layers extract a single-channel 2D latent spatial representation from the input LR flow field.
	This 2D latent representation is used to obtain the scores of each patch via a maxpooling and a softmax layer. It is also output and concatenated with the original LR image.\label{fig:scorernetwork}}
\end{figure}

\textbf{Scorer.} The LR flow field is first input to the \emph{scorer} network. This is a trainable
network whose goal is to score each patch of the LR image via its 2D spatial
latent representation, 
as illustrated in Figure~\ref{fig:scorernetwork}. This network is inspired
by the work of ~\citeauthor{differentiablepatchselection}~\cite{differentiablepatchselection}
that use a similar network for finding salient patches from the input image for classification.

The \emph{scorer} network
consists of a shallow CNN followed
by a maxpooling layer and a softmax layer. 
The first three convolutional layers
extract an abstract representation of the LR flow field. Their
kernel size is $(5,5)$ and the stride is 1. 
This overlap captures the spatial relationships between and among
the input flow variables -- a $(5,5)$ kernel can capture 
both short and long-range dependencies and a stride of 1 maintains the spatial dimensionality. After the
first three convolutional layers, we apply a single-filter convolutional
layer to squeeze and encapsulate the extracted abstract 
spatial information in a single-channel
image. This image is a 2D latent representation
of the spatial dependencies in the variables of the LR flow field and plays a key role in determining the scores of each patch. 
This single-channel
image is input to the maxpooling layer that splits the domain into ${NP}_x \times {NP}_y = N $ patches --
where ${NP}_x$ is the number of patches in the horizontal
direction, ${NP}_y$ is the number of patches in the vertical 
 direction, and $N$ is the total number 
of patches. The pool size
and the stride are both $(ph,pw)$,
where $ph$ is the height of the patch and $pw$ is the width of the patch. Hence,
each value in this image represents the non-normalized score of each patch. The softmax
layer normalizes these scores
to a $0-1$ scaled probability distribution. 
The \emph{scorer} network's output is twofold:
the scaled scores and the 2D latent representation.

Before passing the scores to the \emph{ranker}, 
we concatenate the 2D spatial latent representation with the LR flow field. This is motivated by two reasons. 
First, the latent representation already contains
spatial correlations from the LR flow field. Second, it allows to dynamically change
the scores of the patches since the \emph{scorer's} weights are propagated during the backward pass of the training process.
Hence, the LR flow field becomes a five-channel image (in Figure~\ref{fig:amrnetdnn}, 
$PC=5$). Finally, we pass the scores from the \emph{scorer}
to the \emph{ranker}.

\textbf{Ranker}. This is a non-trainable module
that tracks the score and the ID of each patch.
The \emph{ranker} locates each patch in the new
five-channel LR flow field, isolates it from the rest
of the image, and places it in a bin
according to its score.
We refer to this process as \emph{binning},
and it is illustrated in Figure~\ref{fig:amrnetdnn}.
Binning consists of splitting the $0-1$ range of values
of the scores
into $b$ bins uniformly.\footnote{Another approach is to bin non-uniformly,
which would change each patch's
placement and their final resolution, giving more or less
importance to higher-scored patches. However, in this paper
we only explore uniform binning.} For instance, if $b=2$, the first
bin consists of patches with scores between $0-0.5$,
and the second bin with scores $0.5-1$. The \emph{ranker} plays
a significant role in determining the final resolution of each patch:
training consists of mapping the highest-scored patches
to the highest target resolution,
and as scores gradually decrease, so does the target resolution of the patch.
The patches with the lowest scores remain low-resolution. 

After the binning finishes, we perform two additional
steps before passing the patches to the \emph{decoder}. 
First, we upsample (refine) each patch to its target resolution
using bicubic interpolation, as seen
in Figure~\ref{fig:amrnetdnn}. Then, we concatenate the 2D coordinates to each patch needed to compute the gradients of the flow variables using automatic differentiation~\cite{AD}.
This leads to a seven-channel image (in Figure~\ref{fig:amrnetdnn}, $PC+2$).
Once we upsample and concatenate the coordinates to the patches, it is input to the decoder.

\begin{figure}[htbp]
\centering
\small
\includegraphics[width=1\columnwidth]{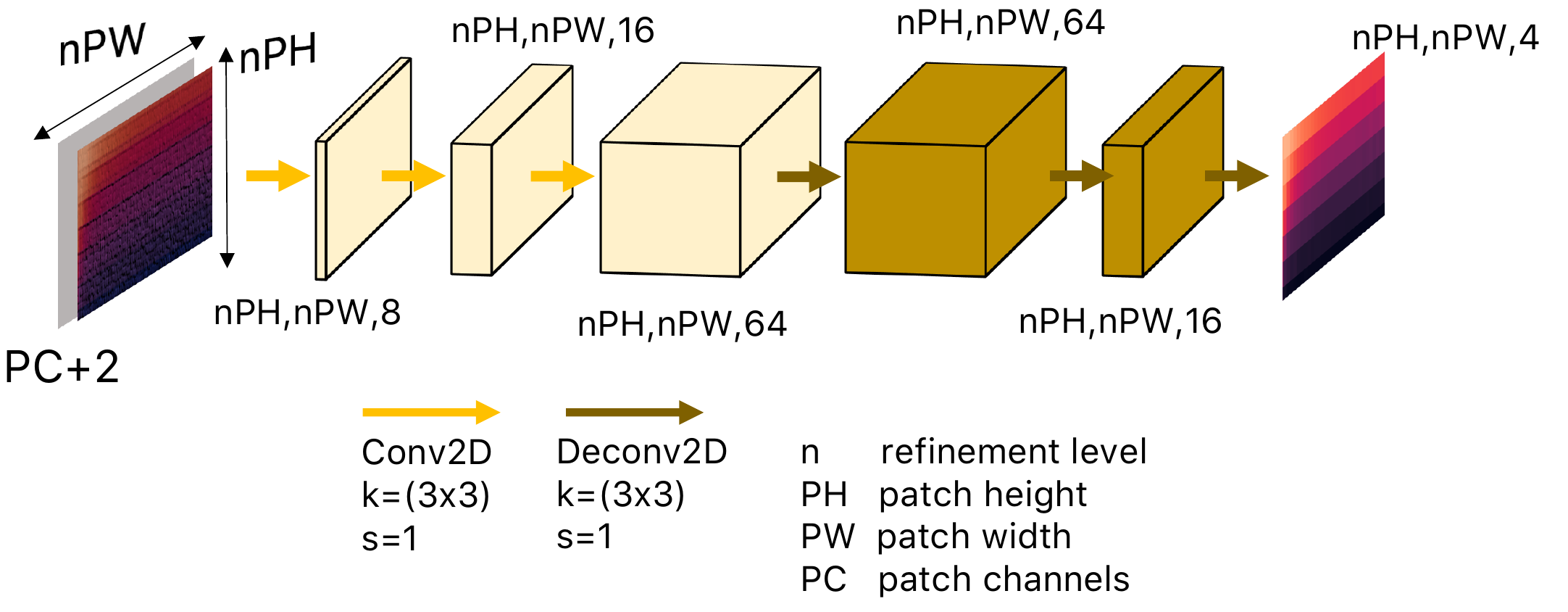}
	\caption{\small The \emph{decoder} network. The input to the decoder is
	the intermediate patch representation concatenated with the
	2D coordinates at its target resolution. This network consists of 3 convolutional layers followed by 3 deconvolutional layers that reconstruct the final values of the patch at its target resolution. The output is, therefore, a four-channel
	image, where each channel is the value of a flow variable.\label{fig:decoder}}

\end{figure}

\textbf{Decoder}. The goal of this trainable network illustrated in Figure~\ref{fig:decoder} is to reconstruct the high-resolution solution of each patch. Each patch in each bin passes through the same decoder, which is shared among resolutions. Note that the patches placed in the low-resolution
bin also passes through the decoder.
The choice of a shared decoder is motivated by two reasons. 
First, we have a smaller number of learnable parameters compared to a separate decoder for each resolution.
Thus, we stress \name's ability to recover different resolutions
for different patches with a lower computational cost. Second,
the low-resolution patches have not been upsampled and the decoder can extract the true spatial correlations
between the flow variables and coordinates in those patches. We expect
this to help in recovering the true values of the high-resolution patches.

This network is a 6-layer network; 3 convolutional layers
followed by 3 deconvolutional layers. The number of filters is
8, 16, 64, 64, 16, 4; the kernel size is $(3,3)$, and the stride is 1.
The choice of this architecture is twofold. First,
recent works have successfully leveraged similar architectures for flow super-resolution~\cite{denoisingsuperresolution,surfnet}. 
Second, the convolutional layers
aim at extracting a deep, abstract representation used by the deconvolution
layers to reconstruct the high-resolution output. We use a stride of 1 to not lose any spatial information. 

The decoder's output consists of a list of patches.
Each patch is a four-channel image, and each channel
represents the values of the four flow variables ($U$, $V$,
$p$, and $\tilde{\nu}$) at steady-state 
at their new spatial resolution. Each patch in the list
can have a different spatial resolution. This list is then passed to the loss function.

\subsection{Loss function}
\label{sec:amrnet:loss}
Recent DL-based super-resolution approaches have leveraged
data-only loss functions~\cite{surfnet,JFMsuperresolution2021},
where CFD simulations generate the ground truth labels.
Other approaches use a combination of data and PDE residual loss function~\cite{meshfreeflownet,denoisingsuperresolution},
where the governing equations are imposed in the loss function. 
We adhere to the latter practice for a couple of reasons. 
First, it is physically and numerically inconsistent to merge CFD-originated
data from different spatial resolutions. 
Second, we do not train with ground truth data from AMR solvers because 
that would make the network learn the solver's heuristics
that have a high degree of user intervention.
The goal is for \name\ to make its own refining decisions
to obtain a DL-based model for AMR. 
Equation~\ref{eq:loss} shows our loss function.

\begin{equation}
	\small
	\mathbb{L} = \frac{1}{np\cdot nc \cdot fv}\sum_{i=1}^{np} \sum_{j=1}^{nc} \sum_{k=1}^{fv}\norm{\hat{y}_{ijk} - y_{ijk}}^2 + \frac{\lambda}{NC \cdot ne} \sum_{i=1}^{NC} \sum_{j=1}^{ne} \norm{P_{j}(i)}^2
	\label{eq:loss}
\end{equation}

Our loss function $\mathbb{L}$ is formed by two parts.
The first term in the right hand side of Equation~\ref{eq:loss}
is the data loss. We take the mean squared error (MSE)
of the prediction of each flow variable ($fv$)
at each cell ($nc$) of the low-resolution patches ($np$) only
with the ground truth data obtained with the physics solver.
The second term in the right hand side is the 
L2 norm of the residual of each PDE ($ne$)
for all cells ($NC$) of the output image, belonging
to either low or high resolution patches. We impose
the continuity equation and the two conservation
of momentum equations (hence, $ne=3$), described in
Equations~\eqref{eq:continuity} and~\eqref{eq:momentum}.
The gradients of the variables are computed through automatic
differentiation~\cite{AD}.
To constrain the PDE
residual of the high-resolution patches,
we downsample them using bicubic interpolation
to the lowest resolution and match the ground truth data in the downsampled space~\cite{denoisingsuperresolution}.
With this semi-supervised learning formulation, we avoid HR labels, an advantage over
SOTA super-resolution methods that require expensive
HR training data~\cite{meshfreeflownet,fourieroperator,neuraloperator,graphkernelnetwork}.

\subsection{End-to-end framework}
\label{sec:amrnet:amrnetvsamrsolvers}
Once the network is trained and calibrated, we use it to
predict the non-uniform high-resolution
flow field of a new problem. However, this prediction
has an approximation error and might not satisfy 
the same convergence constraints
as traditional physics solvers with 
PDE residual values close to the machine round-off errors,
which is critical for many practitioners.
We correct this by refining the DNN's prediction using the physics solver~\cite{maulik,cfdnet,surfnet}.

\begin{figure*}[htbp]
\centering
\small
\includegraphics[width=0.9\textwidth]{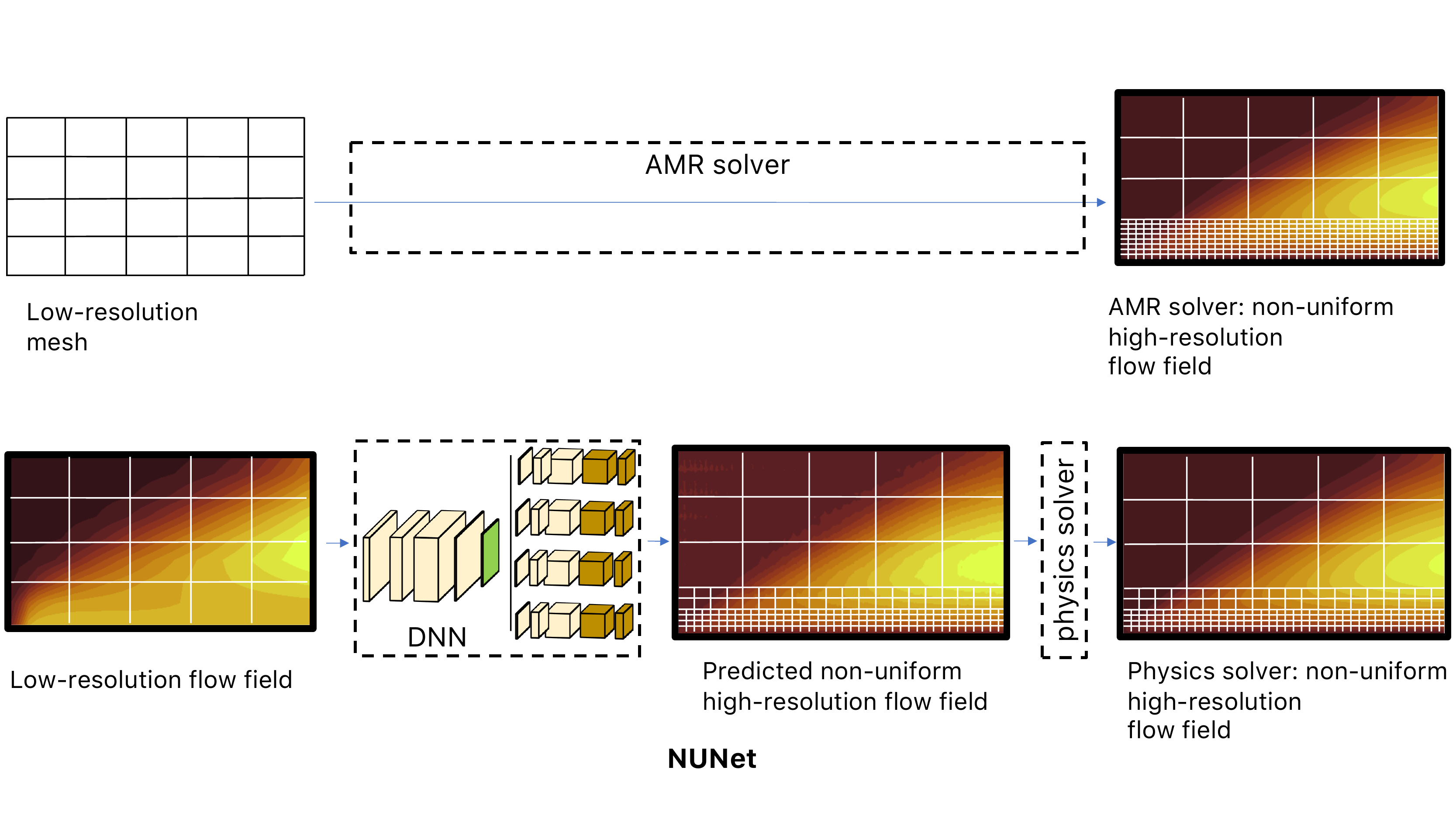}
	\caption{\small Top: Traditional AMR solver simulation. Bottom:
	\name\ framework. After performing non-uniform super-resolution
	with the DNN, we feed the output field into the physics solver,
	which takes the inferred solution to convergence.\label{fig:amrnet}}

\end{figure*}

Figure~\ref{fig:amrnet} illustrates \name's 
end-to-end framework compared with
the traditional AMR solver. In \name, 
the LR flow field is input to a DNN, whose 
architecture is detailed in Section~\ref{sec:amrnet:dnn}.
After inference, we feed the DNN's non-uniform output
into the physics solver, impose the boundary conditions
that well-pose the problem, and let the physics solver
drive this inference to convergence. 
Note that the physics solver does not do
any further refinement or coarsening. The final
discretization is an output of the DNN.

As a result, we obtain a solution that satisfies the same convergence
constraints as traditional numerical methods. Since we anticipate the DNN's inference to be "close" to the final solution, convergence is accelerated. Section~\ref{sec:res:amrnetvsamrsolver} empirically evaluates the performance of \name\ against both classical AMR solvers and SOTA DL models.

\section{Experiment Setup}
In this section, we present the methodology to train
and evaluate \name. We first describe
the governing equations then, present
the dataset
to train and validate \name\ along with the training/testing setup, parameters, and results.
Finally, we describe the physics solver
and the traditional AMR heuristics used for comparison.

\subsection{Background}
\label{sec:exp:background}

We use steady incompressible Reynolds-Averaged
Navier-Stokes problems for our non-uniform super-resolution task.
The RANS equations describe turbulent flows as follows: 

\vspace{-1em}
\begin{align}
  \frac{\partial \bar{U}_i}{\partial x_i} &= 0\label{eq:continuity}\\ \bar{U}_j
  \frac{\partial \bar{U}_i}{\partial x_j} &= \frac{\partial}{\partial
  x_j}\left[ -\left(\bar{p}\right)\delta_{ij}
  +\left(\nu+\nu_t\right)\left(\frac{\partial \bar{U}_i}{\partial x_j}
  + \frac{\partial \bar{U}_j}{\partial x_i}\right)\right]\label{eq:momentum}
\end{align} 

where $\bar{U}$ is the mean velocity,
$\bar{p}$ is the kinematic mean pressure, $\nu$ is the fluid
laminar viscosity, and $\nu_t$ is the eddy viscosity. Equation~\ref{eq:momentum}
is a vector equation that yields two scalar equations - one for each spatial direction. The RANS equations are a time-averaged solution to the incompressible Navier-Stokes equations but
result in a non-closed equation. We close the equation with Boussinesq's
approximation~\cite{pope} by relating the Reynolds stresses with the mean flow: we model the Reynolds stresses with the eddy viscosity and the main rate of the strain tensor. 
We then model the eddy visccosity with the Spalart-Allmaras (SA) one-equation model, that provides
a transport equation to compute the eddy viscosity~\cite{SpalartAllmaras}, described in Equation~\ref{eq:SA}

\begin{equation}
\bar{U}_i \frac{\partial \tilde{\nu}}{\partial x_i } = C_{b1} \left(1-f_{t2}\right) \tilde{S} \tilde{\nu} - \left[C_{w1}f_w - \frac{C_{b1}}{\kappa^2} f_{t2} \right] \left(\frac{\tilde{\nu}}{d}^2\right) \\ + \frac{1}{\sigma} \left[\frac{\partial}{\partial x_i} \left( \left(\nu+\tilde{\nu} \right) \frac{\partial \tilde{\nu}}{\partial x_i}\right) +C_{b2} \frac{\partial \tilde{\nu}}{\partial x_j}\frac{\partial \tilde{\nu}}{\partial x_j} \right]
\label{eq:SA}
\end{equation}

From Equation~\ref{eq:SA} we can compute the eddy viscosity from $\tilde{\nu}$ as $\nu_t = \tilde{\nu} f_{v1}$. These equations represent the most popular implementation of the Spalart-Allmaras model.
The terms $f_{v1}$, $\tilde{S}$, and $f_{t2}$ are model-specific 
and contain, for instance, first order flow features (magnitude of the vorticity). 
$C_{b1}$, $C_{w1}$, $C_{b2}$, $\kappa$, and $\sigma$ are constants specific to the model, 
found experimentally. $d$ is the closest distance to a solid surface. 
We choose the RANS equations with the Spalart-Allmaras turbulence model because 
(1) it is a one-equation model and, therefore, computationally convenient, 
and (2) it has been widely explored in all our benchmark cases (see Section~\ref{sec:experiments:subsec:dataset}), 
including aerospace applications for which SA was designed. 
If we can mimic the AMR behavior with SA,  we are optimistic that our approach can be applied to many RANS turbulence models. 
The constants of the model are those in its original reference~\cite{SpalartAllmaras}.

\subsection{Dataset Overview and Flow Description}
\label{sec:experiments:subsec:dataset}
We gather low-resolution data from three well-known
canonical flows for training the DNN in Figure~\ref{fig:amrnetdnn}.
The resolution for this dataset is $64\times256$,
since it is a common resolution for low-resolution solutions
for all our training cases~\cite{pope}.

\textbf{Turbulent flow in a channel}.
2D channel flow has been widely studied in the literature~\cite{moin}. 
A common strategy to evaluate channel flow
is to vary the input velocity to the channel. This is
the same as varying the Reynolds
\footnote{The Reynolds number, or \Rey, is a non-dimensional coefficient that quantifies the flow conditions of the problem} 
number since \Rey\ $=\frac{UL}{\nu}$, where $U$ is
the input velocity to the channel, $L$ is the diameter of the channel,
and $\nu$ is the laminar viscosity of the problem. Here, we
adhere to this practice and vary the input velocity to the channel to
collect $10000$ samples.
Specifically, we collect 300 samples from \Rey\ = \num{2e3}
(when turbulent effects start to appear~\cite{pope}) to 
\Rey\ = \num{2.3e3}, and then, 9700 more samples from \Rey\ = \num{2.7e3}
to \Rey\ = \num{1.35e4}. We leave \Rey\ = \num{2.5e3} and \Rey\ = {1.5e4} as 
our test cases. Section~\ref{sec:results} presents a more in-depth discussion of the selection of the 
test cases. 
The physical domain of the channel flow is a diameter of 0.1 meters and a length of 6 meters so we find fully developed flow. 
The inlet is at the left and the outlet at the right. 
The top and the bottom are both walls and hence have the no-slip boundary condition. 
The velocity boundary conditions are uniform inlet at the inlet, no-gradient at the outlet, and 0 at the walls. 
The pressure boundary conditions are no-gradient at the inlet, 0 uniform at the outlet, and no-gradient at the walls. 
The modified eddy viscosity boundary conditions are $3\times$(laminar viscosity) at the inlet, 
no-gradient at the outlet, and 0 at the walls. The laminar viscosity is set to \num{1e-4} \si{\meter^2\per\second} .

\textbf{Turbulent flow over a flat plate}. Flat plate
is also a canonical flow, part of the wall-bounded
flows family, used to study the boundary layer in both
laminar and turbulent conditions~\cite{pope}. By varying
the incoming velocity we collect 10000 samples.
For flat plate, incompressible 
turbulent effects do not appear~\cite{pope} up until \Rey\ = \num{1.35e5}
and scale up to \Rey\ = \num{5e6}. We collect
2000 samples from \Rey\ = \num{1.35e5} to \Rey\ = \num{2e5} and another 8000 additional samples from \Rey\ = {3e5} until \Rey\ = {1.1e6}. We leave \Rey\ = \num{2.5e5} and \Rey\ = \num{1.35e6} as test cases.
The physical domain of the flat plate case is a height of 0.2 meters and a length of 10 meters, 
as found in different benchmarks. 
The boundaries are a wall at the bottom (the flat plate), symmetry at the top, an inlet at the left, and an outlet at the right. 
The velocity boundary conditions are uniform inlet at the inlet, no-slip condition at the bottom wall, 
no-gradient both at the outlet. 
The pressure boundary condition is no-gradient at the inlet and bottom wall, and 0 at the outlet. 
The modified eddy viscosity boundary conditions are $3\times$(laminar viscosity) at the inlet, 
0 at the bottom wall, and no-gradient at the outlet. 
The laminar viscosity is set to \num{1e-4} \si{\meter^2\per\second}.

\begin{figure}[htbp]
\centering
\includegraphics[width=1\columnwidth]{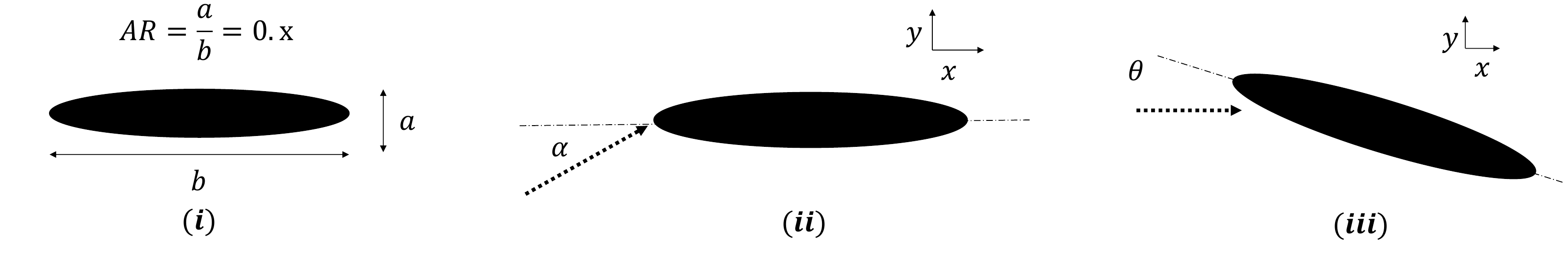}
	\caption{\small Sketch of the (i) ellipse aspect ratio, (ii) angle of attack $\alpha$, and (iii) pitch angle $\theta$.}
	
\label{fig:ellipse_pitch_attack}
\end{figure}

\textbf{Turbulent flow around ellipses}. External
aerodynamics simulations are relevant for aerospace industrial applications. 
We gather low-resolution solutions
from flow around ellipses. In real scenarios,
different geometries at a variety
of flow conditions are explored. Our training data
consists of 10000 samples of flow around different ellipses
at different flow conditions. Figure~\ref{fig:ellipse_pitch_attack}
shows these configurations.

The training data is obtained by changing the aspect ratio of the ellipses:
$0.05, 0.07, 0.09, 0.1, 0.15, 0.2, 0.25, 0.35, 0.55$, and $0.75$.
Each of these ellipses is simulated under 5 different flow conditions
by changing randomly the angle of attack $\alpha$ and the
pitching angle $\theta$ between $-2$ and $6$ degrees. We collect all of these configurations at 200 different \Rey\ numbers
between \num{5e4} and \num{9e4}. We select
flow around a cylinder at \Rey\ = \num{1e5}, flow around a
symmetric airfoil (NACA0012) at \Rey\ = \num{2.5e4}, and flow around a non-symmetric airfoil (NACA1412) at \Rey\ = \num{2.5e4}.
The physical domain of the ellipse/cylinder/airfoil cases consists of a solid body of chord ( c ) 1 meter, 
and the far-field limit is located 30c from the tip and tail of the solid body (O-grid type of mesh). 
The velocity boundary conditions are no-slip at the wall and uniform velocity at the far-field; 
the pressure boundary conditions are no-gradient at the wall and uniform 0 in the far-field; 
the modified eddy viscosity boundary conditions are 0 at the wall and uniform $3\times$(laminar viscosity) in the far-field, 
where the laminar viscosity is \num{1e-4} \si{\meter^2\per\second} uniform in the entire domain. 

The total training set size is composed of $30000$ samples,
$10000$ from each canonical flow. From this training dataset,
$27000$ samples are used for training the DNN and
$3000$ samples are used for validation. 

\subsection{Training and Testing Setup}
\label{sec:exp:training}
The methodology described in Section~\ref{sec:amrnet}
allows for multiple combinations of  parameters in addition to those
in neural network optimization.
For instance, we can select different patch sizes or different number of bins
(and therefore, different number of target resolutions). In this section,
we explain our training design choices. 

First, \name's DNN's convolutional kernels of size $5\times5$ and $3\times3$
require a minimum input image size to extract relevant information.
Therefore, we fix our patch sizes at $ph\times pw = 16\times16$, which leads to 
$N=64$ total number of patches for each train/validation/test sample. 
Larger patch sizes (for instance, $32\times 32$) do not offer
enough granularity to make critical distinctions
between regions of the flow. 
Second, we choose the number of bins
$b=4$, and hence four different target resolutions. Each target
resolution refines the original low-resolution patch
by $4^{(n)} \times$, where $n=0,1,2,3$. We choose
$b=4$ because not more than 4 levels of refinement is an extended practice
in the AMR literature~\cite{bergerAMRhyperbolic, featurebasedAMR}
to avoid tiny computational cells. This also allows us to compare \name\ with SOTA approaches that attempt $64\times$ 
super-resolutions. 
The patch size and the number of bins are the same at test time
and during the evaluation of the results. 

We implement the DNN
using the Tensorflow 2.4 backend,
and perform distributed training on four Tesla V100 GPUs connected with PCIe.
After training the network with a batch size of $8$, a learning rate
of $1e-4$, no specific initialization, and using the Adam optimizer~\cite{adam} for 350 epochs,
the training and validation data and PDE residual loss for all equations reach a MSE of 9e-6.
Note that the training of \name's DNN is done on the Tesla V100 GPUs. However,
\name\ is implemented entirely on the CPU for a fair comparison with the AMR solver,
which only supports CPU. Hence, both \name\ and the AMR solver are executed on the CPU.

\subsection{Physics solver and AMR solver}
\label{sec:experiment:subsec:solvers}
Once the network is calibrated,
it is used for inference. Recall
that the DNN's output is input
into the physics solver to drive
the flow field from inference to convergence (see Section~\ref{sec:amrnet:amrnetvsamrsolvers}). 
We use OpenFOAM's \texttt{pimpleFoam}~\cite{openfoam,pimplefoam} 
solver as the physics solver in this paper.
For the pressure, we use the \texttt{GAMG} solver,
with a relative tolerance of 0.1,
and the \texttt{GaussSeidel} smoother from OpenFOAM. 
For the velocity and the eddy viscosity, we use the \texttt{smoothSolver}
with a relative tolerance of 0.1. The number of
outer and inner correctors are set to 1, and the time scheme
is set to steady-state.

As for the AMR solver to compare \name\ with, 
we use the \texttt{dynamicMeshRefine}
utility in OpenFOAM. This utility performs AMR as long
as it is used together with \texttt{pimpleFoam} solver. The
combination of the \texttt{pimpleFoam} solver with the\\
\texttt{dynamicMeshRefine} utility forms the AMR solver
used in this paper. 
This solver is a feature-based
AMR solver, which is the most popular method in the literature~\cite{AMReX2021}.
Therefore, it requires user intervention: for all
of the test cases, we set the AMR solver to refine
those areas where the gradients of the eddy viscosity are the highest,
and the maximum level of refinement is set to 4.
This heuristic is popular and works well for 
a wide range of problems, including our test problems~\cite{pope}. 

\textbf{Architecture and Libraries}. 
All the OpenFOAM simulations are run in parallel 
on a dual-socket Intel Xeon Gold 6148 using double precision due to the lack of GPU support. 
Each socket has 20 cores, for a total of 40 cores. 
We use the OpenMPI implementation of MPI integrated with OpenFOAM v8 that is optimized for shared-memory communication. 
The grid domain is decomposed into 40 partitions 
using the integrated Scotch partitioner and each partition 
is assigned to 1 MPI process that is pinned to a single core. 
We set the \texttt{numactl -localalloc} flag to bind each MPI process to its local memory.

\section{Results and Discussion}
\label{sec:results}
\begin{figure}[htbp]
\centering
\small
\includegraphics[width=1\columnwidth]{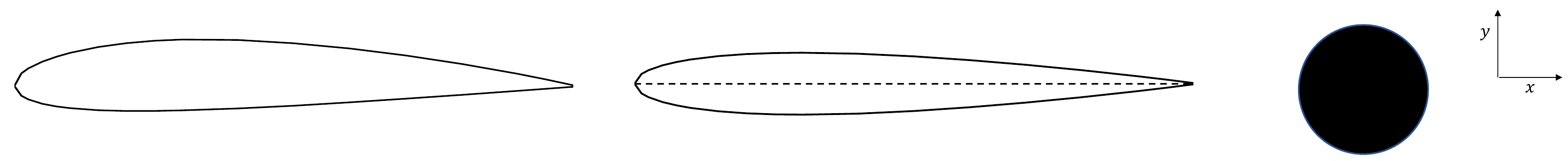}
	\caption{\small Non-symmetric NACA1412 airfoil (left), symmetric NACA0012 airfoil (center), and cylinder (right) as test geometries. These test cases stress the generalization capacity of \name\ on unseen-during-training geometries.}
\label{fig:test_airfoils}
\end{figure}

After training and validating the network, we evaluate \name's
ability for non-uniform super-resolution on two different use cases,
as outlined below:
\begin{enumerate}
	\item[I] \textbf{Same geometry, different boundary conditions}. 
		We use \name\ to refine the LR solution of flow on a geometry observed during
		training but at a different boundary condition. Here, our test flows configurations are
		channel flow and flat plate on interpolated (int) and extrapolated  (ext) boundary conditions. 
		For the former, we test on  \Rey\ = \num{2.5e3} (int) and \Rey\ = \num{15e3} (ext). For the latter,
		we test on \Rey\ = \num{2.5e5} (int) and \Rey\ = \num{1.35e6} (ext).
	\item[II] \textbf{Different geometry, different boundary conditions}. We stress
		the generalization capacity of \name\ by finding the non-uniform high-resolution
		solution of flow around geometries unseen during training. 
		We use the same network to predict the flow around a cylinder
		at \Rey\ = \num{1e5}, the flow around
		a symmetric NACA0012~\cite{propergrid} airfoil at\Rey\ = \num{2.5e4},
		and the non-symmetric NACA1412 airfoil at \Rey\ = \num{2.5e4}, as seen in Figure~\ref{fig:test_airfoils}.

\end{enumerate}

For the described test cases,
we first present the accuracy and correctness
of \name\ by comparing it, both
qualitatively and quantitatively, with
the traditional AMR solver (described in Section~\ref{sec:experiment:subsec:solvers}).
Then, we present the performance analysis
of \name\ by showing (1) its speedup 
over the AMR solver and (2) the improved inference time and memory usage
over the SOTA neural network models that
perform uniform super-resolution. The baseline neural
network used for comparison is described in Section~\ref{sec:results:subsec:performance}.

\subsection{Correctness and Accuracy of \name}
\label{sec:res:amrnetvsamrsolver}
Once the DNN is trained, we build
the \name\ framework and study its
correctness and accuracy by comparing
it with the traditional AMR solver. 
We first conduct a qualitative evaluation by visualizing (a) the refined/unrefined areas 
and (b) the final flow field by the two algorithms. Because of the difference
in their inherent heuristics (\name\ follows
an optimization process containing different
flow cases while the AMR solver follows user-given heuristics
as explained in Section~\ref{sec:experiment:subsec:solvers}),
we do not expect the exact same output. 
However, the qualitative results evaluate whether \name\
can act as an AMR surrogate for multiple flow problems
resulting from a single training.

After, we present a quantitative
comparison between the two using a grid convergence study~\cite{gridrefinementstudies}. Recall that both
\name\ and the AMR solver solve
the same problem.
We impose the same strong-form boundary conditions
in the fluid domain, 
which well-pose the problem and guarantee uniqueness~\cite{chaosinsystems}. 
The only metric that changes between the two is the mesh, and therefore,
they will present different discretization errors. However,
these discretization errors reduce as we increase the resolution of
refinement and global quantities tend to converged solutions~\cite{pope}.
Hence, to evaluate the quality of \name's inferred mesh, we compare 
the solution from both \name's mesh and the AMR solver mesh as we increase
the required levels of refinement. Both meshes are refined
$4^n\times$ gradually, from $n=0$ to $n=3$. Then,
we report the value of specific quantities of interest (QoI) at steady-state. 
The choice of the QoI follows the CFD literature~\cite{pope}.

\subsubsection{Qualitative Results}

\begin{figure}[htbp]
\centering
\small
\includegraphics[width=0.8\columnwidth]{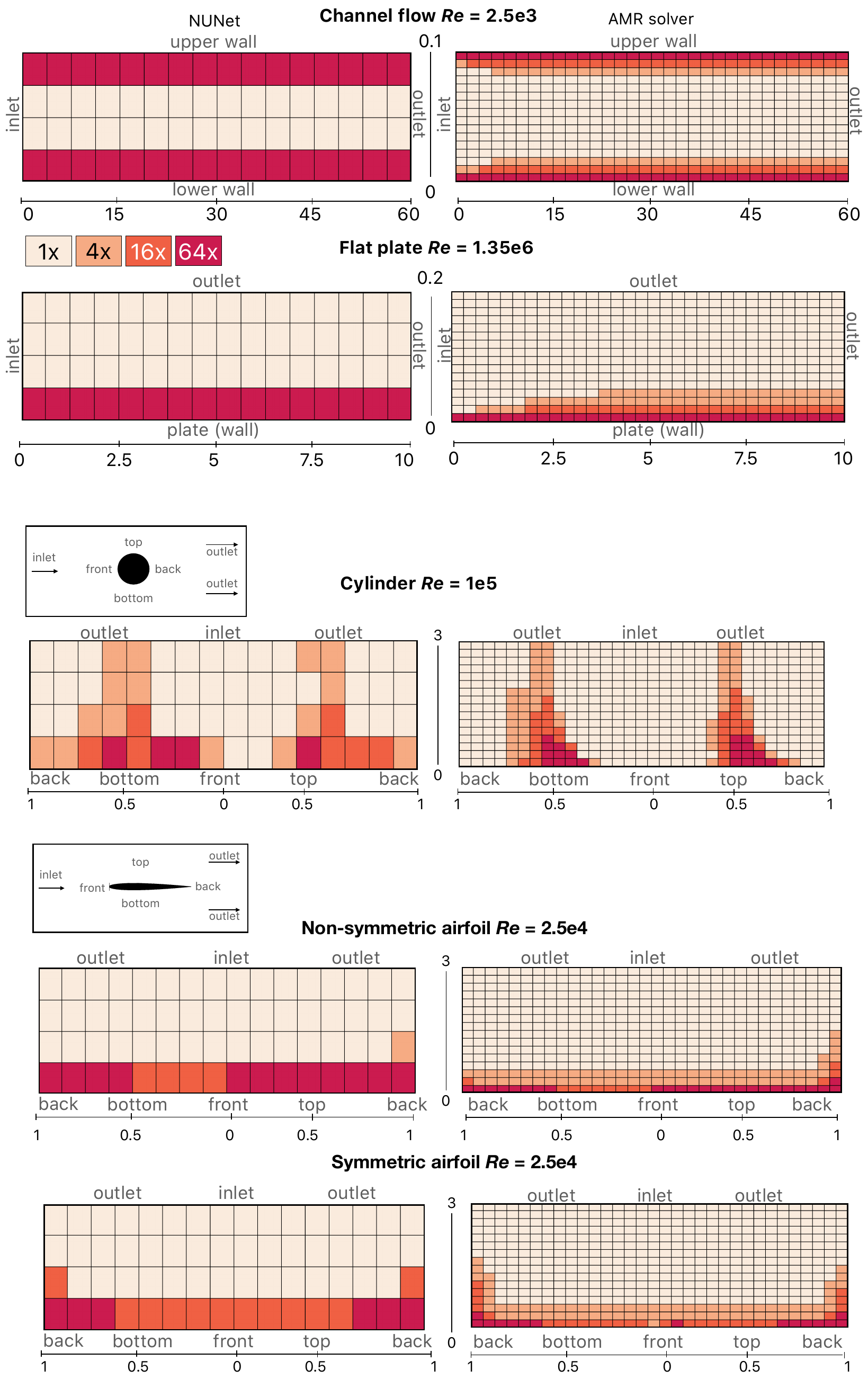}
	\caption{\small Per-patch fluid domain and the level of refinement of each patch
	for all our test cases. First row: channel flow at \Rey\ = \num{2.5e3}. 
	Second row: flat plate at \Rey\ = \num{1.35e6}.
	Third row: cylinder at \Rey\ = \num{1e5}.
	Fourth row: non-symmetric NACA1412 airfoil at \Rey\ = \num{2.5e4}.
	Fifth row: symmetric NACA0012 airfoil at \Rey\ = \num{2.5e4}.
	We compare \name's prediction (left)
	versus AMR solver's output (right). Both axes are in meters.
	\label{fig:patches}}
\end{figure}

Figure~\ref{fig:patches} shows the refined/unrefined
results for five test cases: channel flow at \Rey\ = \num{2.5e3}, 
flat plate at \Rey\ = \num{1.35e6}, cylinder,
and the two airfoil
test cases.
Figure~\ref{fig:patches}
shows \name's predicted mesh (left) and the AMR solver's output mesh (right).
\name\ splits the domain into 64 $16\times16$ patches and we show
the output resolution (with respect to the coarse resolution) of each patch. 
Because the AMR solver allows more granularity as it performs
mesh refinement on a per-cell basis, the domain is divided into smaller ($4\times8$)
patches\footnote{We do not show per-cell refinement as too many cells are
created to offer good visualization. However, $4\times8$ patch sizes
have been found optimal for both gathering cells with equal levels of refinement and visualization quality.}. At the borders of each test case, 
we show the physical boundaries of each case which play a key role in determining the areas where both algorithms refine the mesh.

We make three main observations. 
First, \name\ can distinguish between boundary conditions. 
For the channel flow case (first row in Figure~\ref{fig:patches}),
\name\ refines the fluid areas close to both the upper and the lower wall, 
whereas, for the flat plate case (second row), it refines the areas close
to the wall but leaves the outer regions (outlet/freestream) at low resolution. 
Second,
\name\ respects the symmetry of the problem, 
as we observe in the channel flow case (first row of Figure~\ref{fig:patches}) and in the symmetric airfoil case (fifth row in Figure~\ref{fig:patches}). Third,
\name's fine/coarse regions are in excellent agreement with those of the AMR solver
for the channel flow, flat plate, and airfoil cases.
This agreement in the cylinder case is also notable.
For instance, \name\ refines the region of the flow
from the back of the cylinder to the outlet (\ie
the wake behind the cylinder). However,
we observe some discrepancy in the back-bottom-front-top region. 
The front-bottom-back and front-top-back regions (which refer
to the entire solid boundary of the cylinder) require a higher resolution
from \name. This difference, together with
the channel flow and flat plate results,
indicates that the DNN is refining those areas
with higher values of the gradients for \emph{all}
fluid variables, which take place
in solid wall boundaries~\cite{pope}.
This is opposed to the AMR solver's heuristic,
that focuses only on areas with high gradients of the eddy viscosity. 

During the calibration of the DNN, it is key to balance
both components of the loss function - the data loss and the PDE residual 
loss (described in Section~\ref{sec:amrnet:loss})  
so neither dominates the other. 
In our experiments,
we observe the best predictive results at $\lambda=0.03$, which yields
a balanced contribution of each component of the loss.
The data and PDE residual loss reached a value of \num{9e-6} for both the training and the validation samples.
During training, we scale the value of the variables
between 0 and 1 for learning stability purposes. However,
we can not scale the value of the gradients found by automatic differentiation
because this would result in inconsistent PDE residual loss. These
gradients reach higher absolute values than those of the data,
especially in areas of the flow with higher variability,
and hence get the attention of the MSE loss function. Moreover,
this also allows \name\ to refine the back-outlet area,
where the wake region of the flow after the cylinder meets the freestream (outlet) and we find a
high gradient of the eddy viscosity. The difference in the refining
patterns between the cylinder and the airfoil case is that the former presents flow separation
from the wall boundary that generates a wide wake region, 
whereas in the airfoil case the flow remains attached to the solid. 

\begin{figure}[htbp]
\centering
\small
\includegraphics[width=0.9\columnwidth]{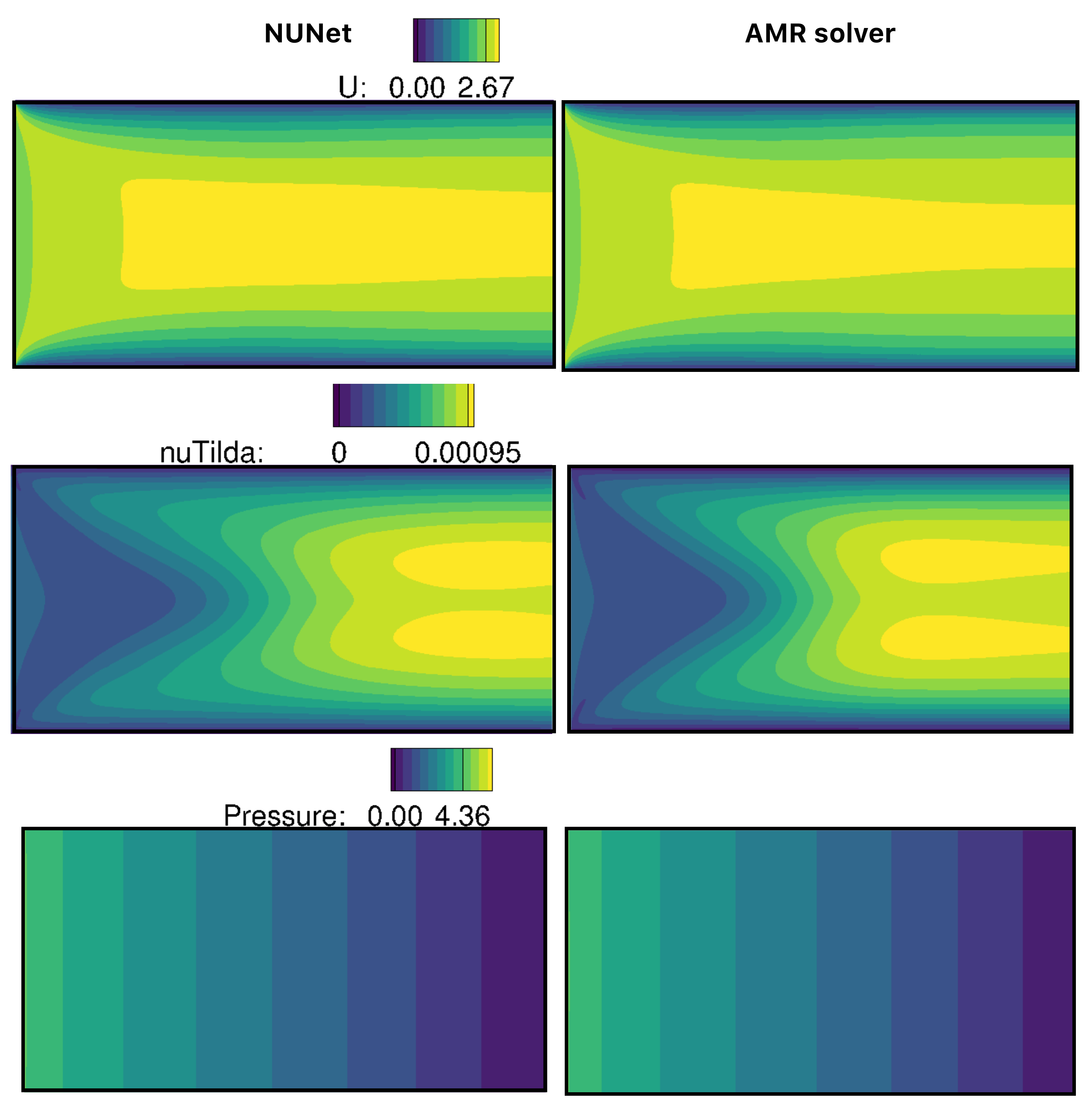}
	\caption{\small Channel flow at \Rey\ = \num{2.5e3}. Velocity in \si{\meter\per\second} (top), kinematic pressure in \si{\meter^2\per\second^2} (bottom), and modified eddy viscosity in \si{\meter^2\per\second} (middle). Comparison between \name's result (left) and the AMR solver result (right) for $b=4$ levels of refinement.\label{fig:variables_cf}}
\end{figure}

\begin{figure}[htbp]
\centering
\small
\includegraphics[width=0.9\columnwidth]{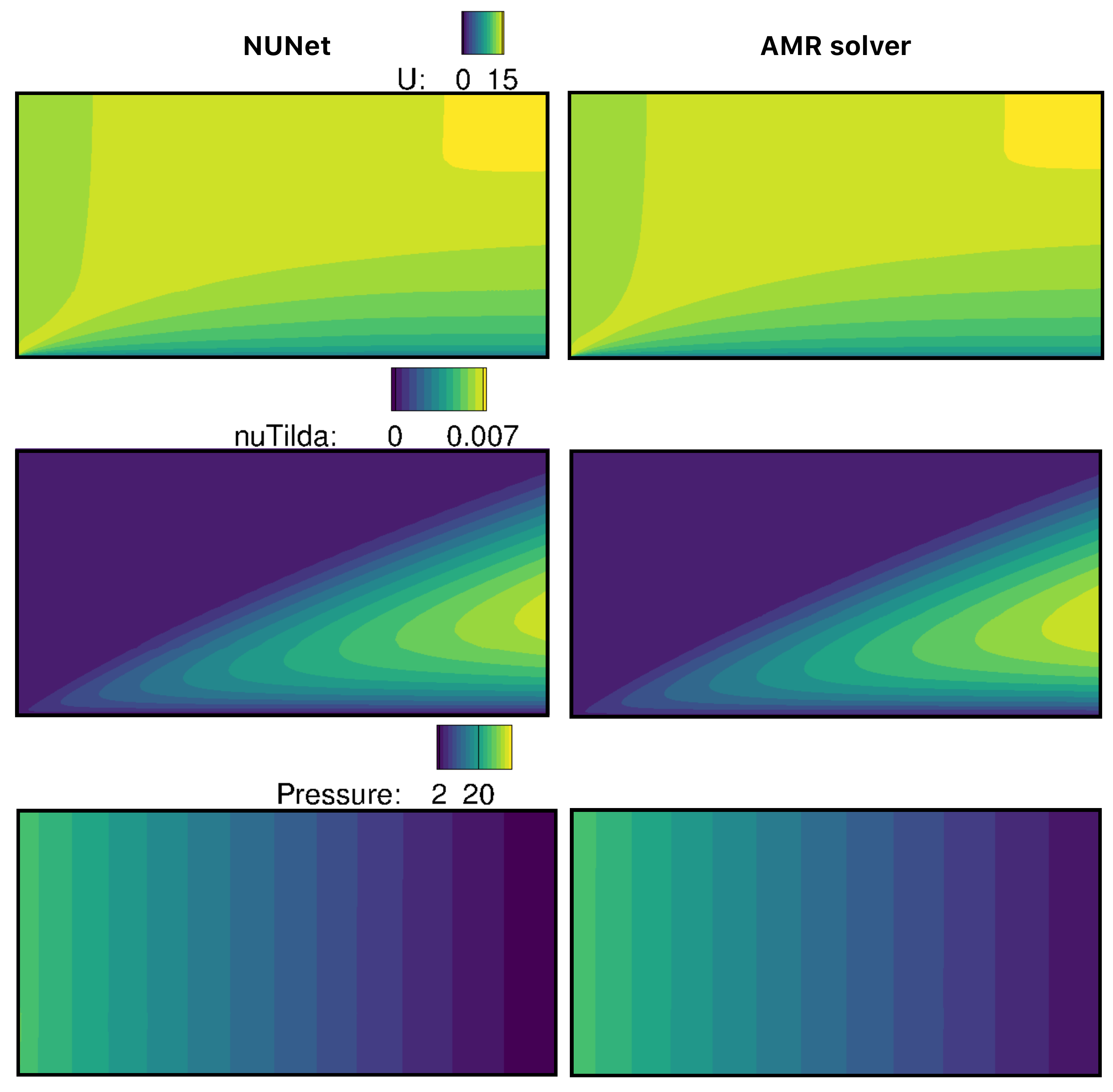}
	\caption{\small Flat plate at \Rey\ = \num{1.35e6}. Velocity in \si{\meter\per\second} (top), kinematic pressure in \si{\meter^2\per\second^2} (bottom), and modified eddy viscosity in \si{\meter^2\per\second} (middle). Comparison between \name's result (left) and the AMR solver result (right) for $b=4$ levels of refinement.\label{fig:variables_fp}}
\end{figure}

\begin{figure}[htbp]
\centering
\small
\includegraphics[width=0.9\columnwidth]{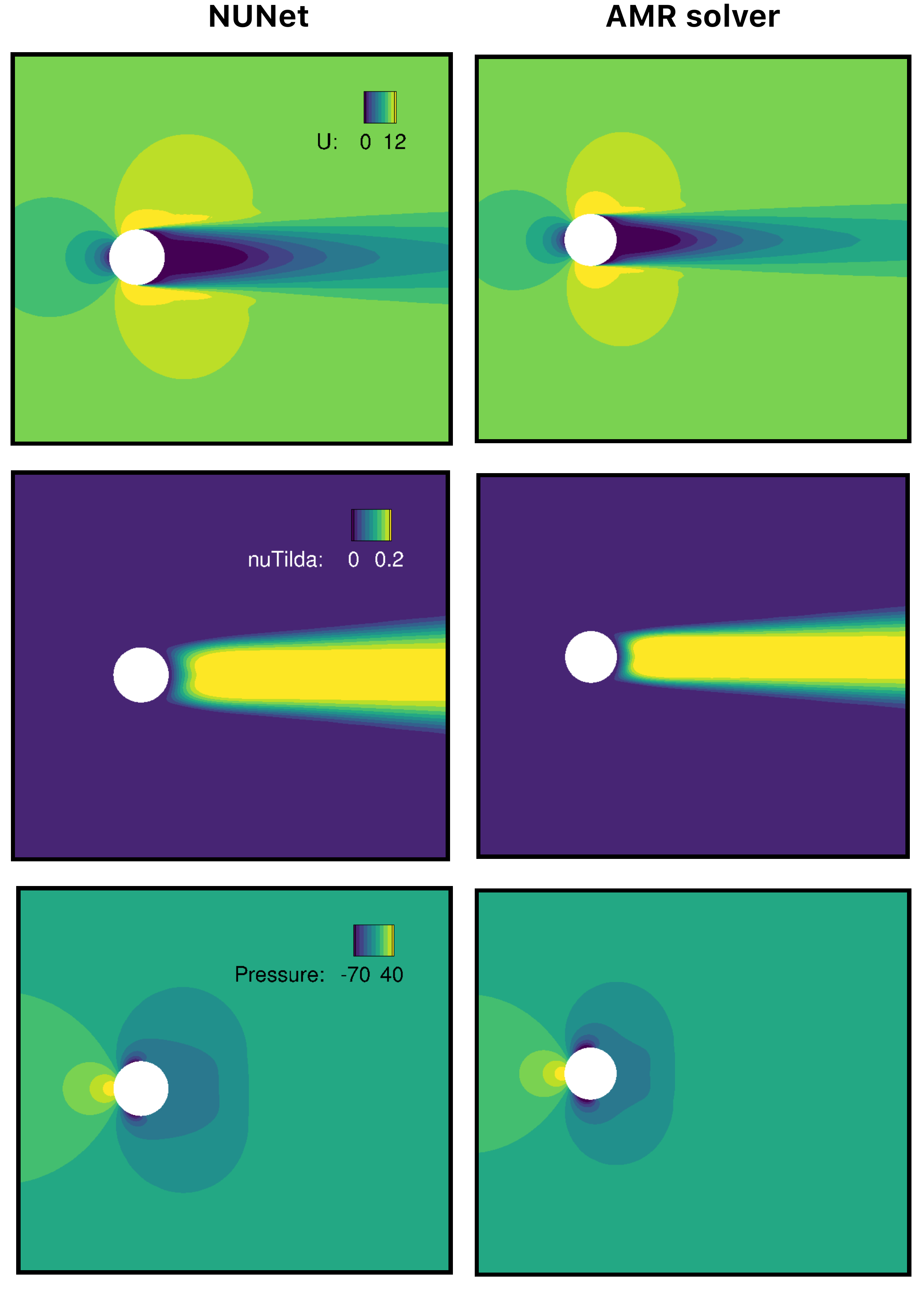}
	\caption{\small Flow around a cylinder at \Rey\ = \num{1e5}. Velocity in \si{\meter\per\second} (top), kinematic pressure in \si{\meter^2\per\second^2} (bottom), and modified eddy viscosity in \si{\meter^2\per\second} (middle). Comparison between \name's result and the AMR solver result for $b=4$ levels of refinement.\label{fig:variables_cylinder}}
\end{figure}

\begin{figure}[htbp]
\centering
\small
\includegraphics[width=0.9\columnwidth]{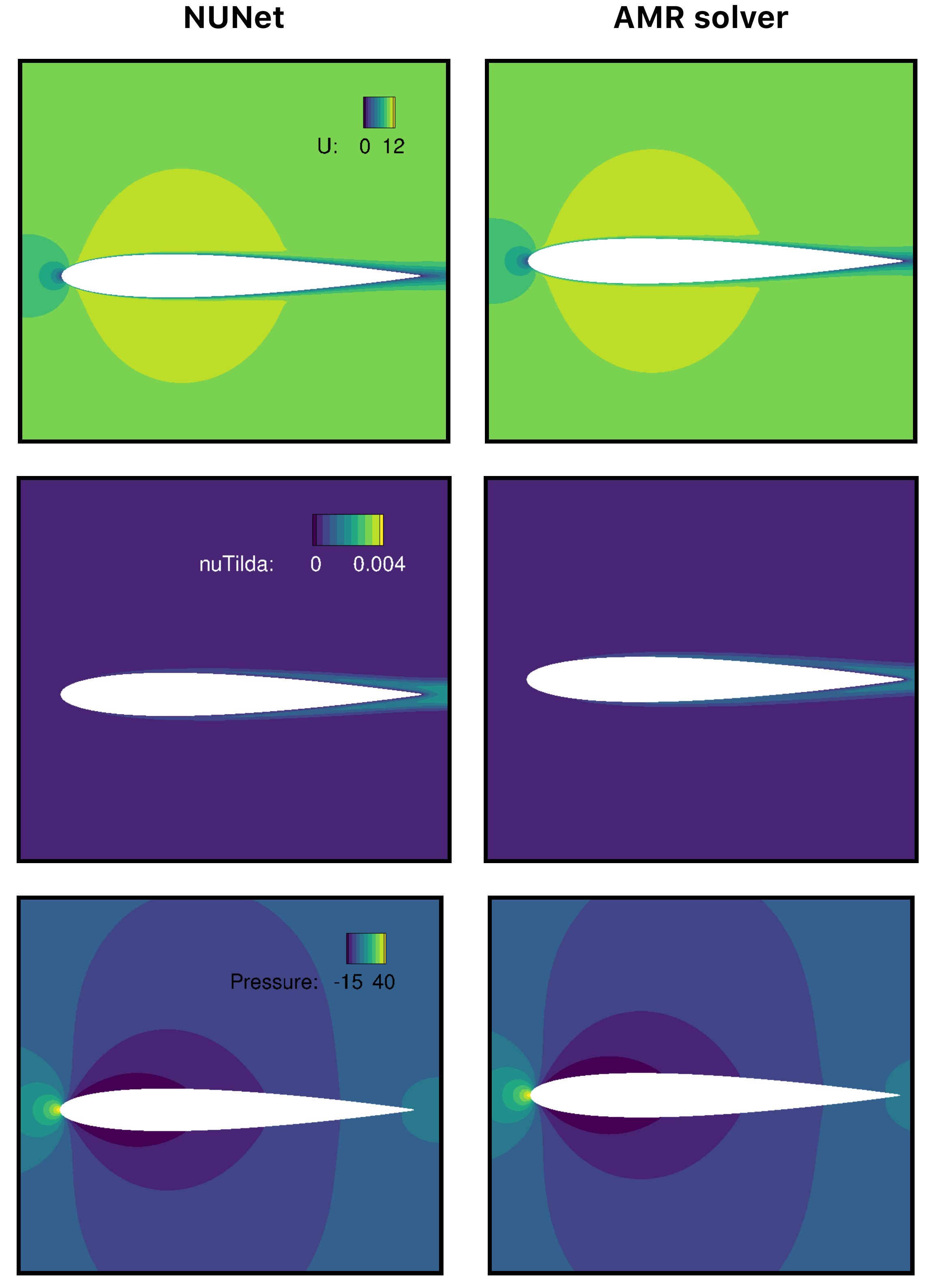}
	\caption{\small Flow around the symmetric NACA0012 airfoil at \Rey\ = \num{2.5e4}. Velocity in \si{\meter\per\second} (top), kinematic pressure in \si{\meter^2\per\second^2} (bottom), and modified eddy viscosity in \si{\meter^2\per\second} (middle). Comparison between \name's result and the AMR solver result for $b=4$ levels of refinement.\label{fig:variables_naca0012}}
\end{figure}

\begin{figure}[htbp]
\centering
\small
\includegraphics[width=0.9\columnwidth]{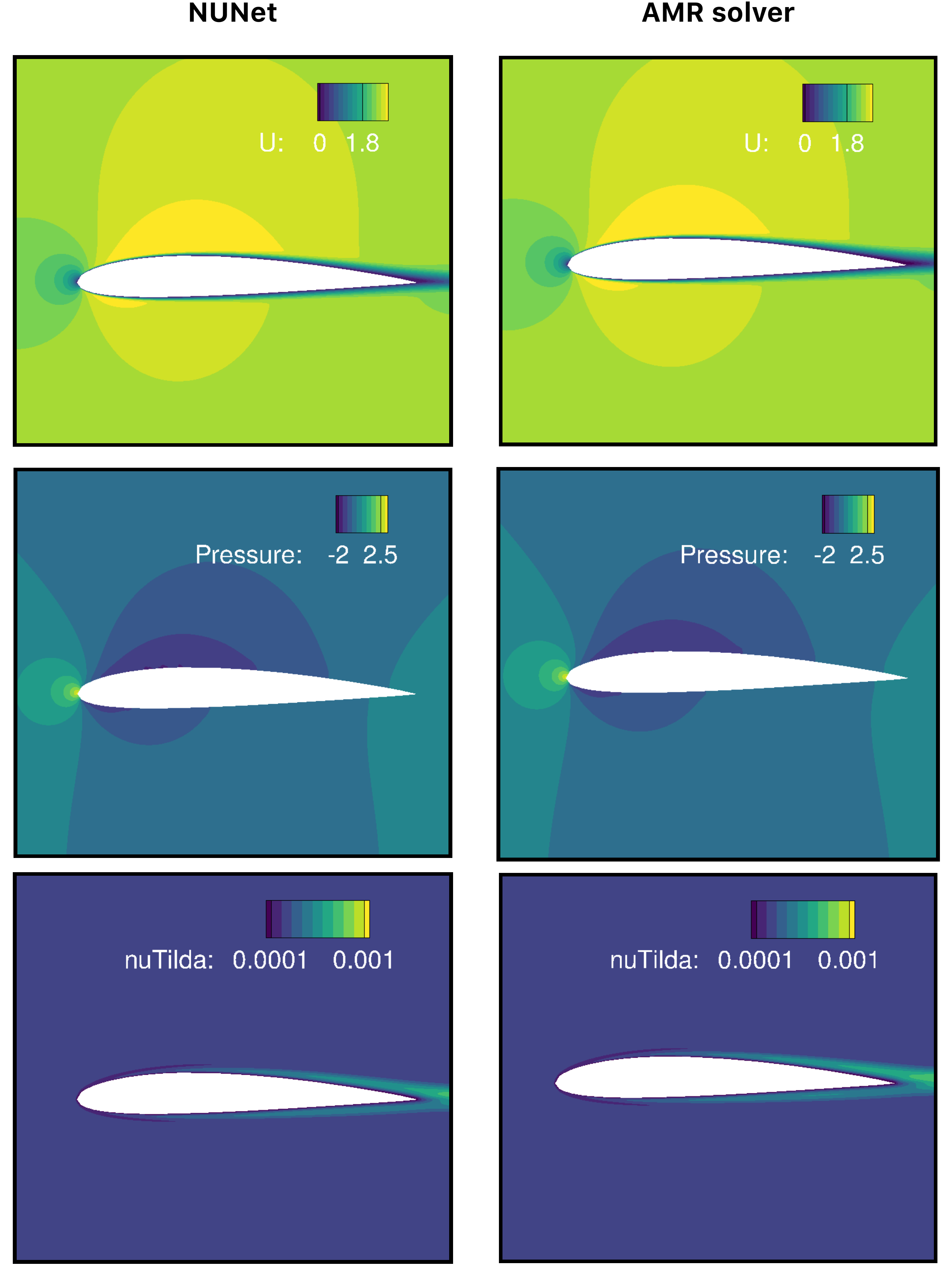}
	\caption{\small Flow around the non-symmetric NACA1412 airfoil at \Rey\ = \num{2.5e4}. Velocity in \si{\meter\per\second} (top), kinematic pressure in \si{\meter^2\per\second^2} (bottom), and modified eddy viscosity in \si{\meter^2\per\second} (middle). Comparison between \name's result and the AMR solver result for $b=4$ levels of refinement.\label{fig:variables_naca1412}}
\end{figure}

Figure~\ref{fig:patches} also shows
that in the channel flow, flat plate, and airfoil test cases, 
the AMR solver reduces the refinement level gradually as we increase the distance from the wall boundary. 
Instead, \name\ infers the maximum level of refinement in the patches
close to the wall and does not show this gradual reduction. This is due to
the maxpooling layer in the design of \name's \emph{scorer} network (see Section~\ref{sec:amrnet:dnn}). Recall that the maxpooling layer chooses
the highest score present in the $16\times 16$ region defined by the patch. Choosing a maxpooling
layer over an average pooling is a desired conservative
approach. 
Since an entire patch shares a resolution in \name, 
it is advantageous to choose 
the highest required resolution even if only few cells within a patch require it 
for accuracy. Figure~\ref{fig:patches}
shows that \name\ and the AMR solver have inherently different heuristics
for mesh refinement/coarsening and do not
produce the same mesh - as expected. However,
both are in excellent agreement
in their steady flow field prediction,
as we qualitatively observe
in Figures~\ref{fig:variables_cf},~\ref{fig:variables_fp},~\ref{fig:variables_cylinder},
~\ref{fig:variables_naca0012}, and~\ref{fig:variables_naca1412}.

\subsubsection{Quantitative results}
\begin{figure*}[htbp]
\centering
\small
\includegraphics[width=1\textwidth]{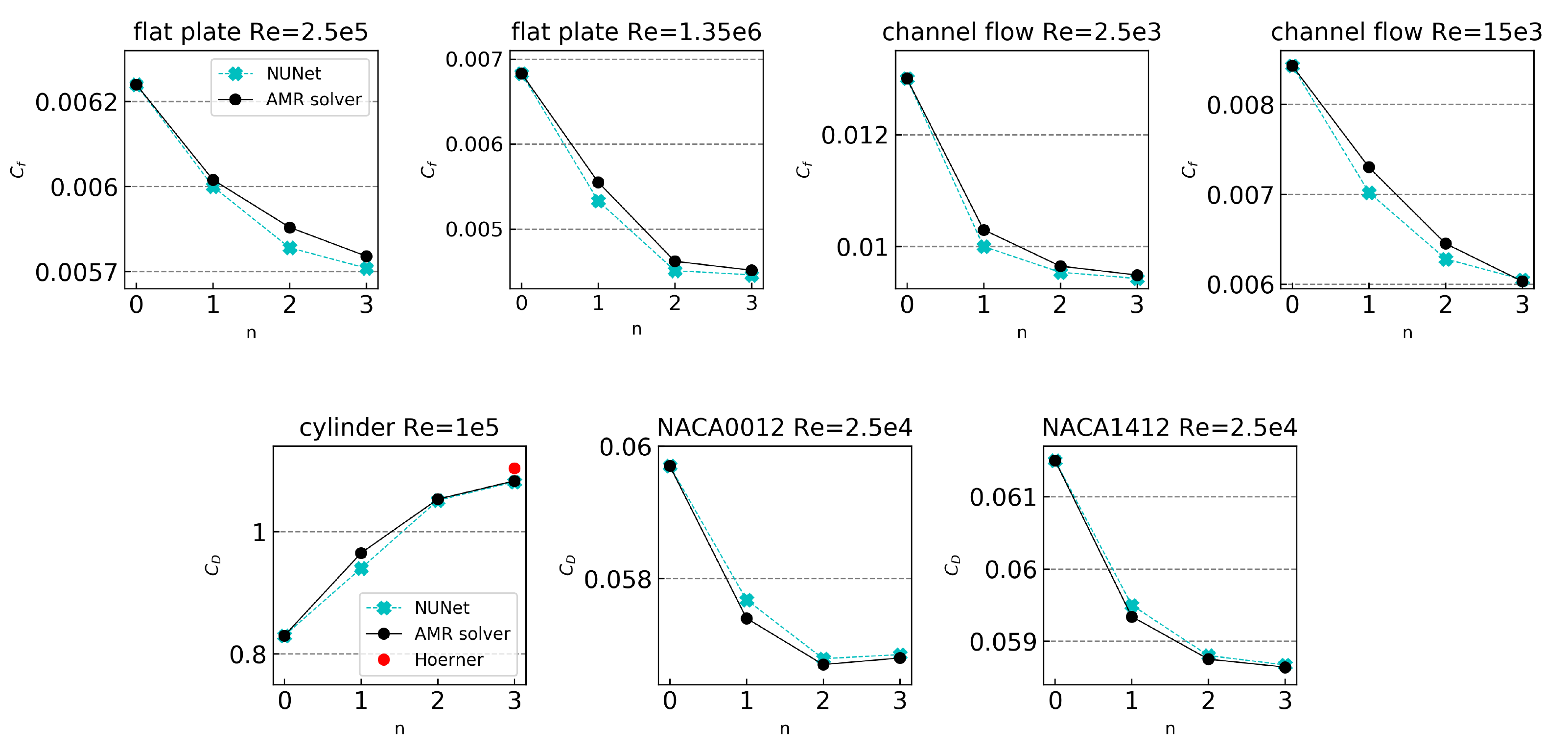}
	\caption{\small Value of the QoI versus refinement level $n$ for \name\ (blue) and the AMR solver (black) for each test case. $C_f$ refers to coefficient of friction, and $C_D$ to coefficient of drag. The red dot is the experimental value for the cylinder case found in~\cite{hoerner}. Both algorithms converge as we increase the mesh refinement level from the original coarse mesh.
	\label{fig:gridconvergence}}
\end{figure*}

We present a quantitative comparison 
between \name\ and the AMR solver using a
grid convergence study.
We report, for the flat plate test cases, the coefficient of friction ($C_f$) 
at $x=0.95L$, where $L$ is the length of the flat plate. For the channel flow test cases,
we also report the $C_f$ on the lower wall at $x=0.95L$. For the cylinder and 
airfoil test cases,
we monitor the coefficient of drag or $C_D$. Figure~\ref{fig:gridconvergence}
shows the value of the QoI for each test case with increasing refinement level $n$. 

We make two main observations from
the plots in Figure~\ref{fig:gridconvergence}. First, we observe a good agreement between
the QoI reported by \name\ and the AMR solver at all levels of refinement. At $n=0$, the value
of the QoI is the same because they start with the same coarse mesh. Second, we observe how
\name's and the AMR solver's reported QoI show a notable
convergence trend after $n=1$. The plot
referring to the cylinder case in Figure~\ref{fig:gridconvergence} shows, in red, the 
experimental value of $C_D$ reported in~\citeauthor{hoerner}~\cite{hoerner},
which is $1.108$, while \name\ reports $1.0835$, a $2.2\%$ deviation,
and the AMR solver reports $1.085$, a $2.1\%$ deviation. These
errors are in line with those in the literature
when comparing experimental results with RANS simulations using the SA model~\cite{bao2011numerical}.

\subsection{Performance Analysis of \name}
\label{sec:results:subsec:performance}
In this section , we evaluate \name's performance.
We first compare
its time-to-convergence (TTC) with the
AMR solver's TTC in 
obtaining the results in Figure~\ref{fig:gridconvergence} at $n=3$. 
Recall that
\name's TTC is the sum of the inference
time and the time the physics solver takes to drive the solution from inference to convergence.
Table~\ref{tab:vsphysicssolver} shows these times
and reports the iterations-to-convergence (ITC)
taken by both the physics solver and the AMR solver. 
For the channel flow case, \name\ achieves a $4.3\times$ speedup for the interpolated
boundary condition case.
For the flat plate,
we obtain a $5.5\times$ (interpolated case) and a $4.7\times$ speedup (extrapolated case) over the AMR solver. \name\
obtains an impressive $3.2\times$ speedup for a flow around a cylinder, 
which is an unseen-during-training geometry. 
The cylinder case is the most challenging
test case for \name\ since accurately
predicting the wake region behind the cylinder (as seen in Figure~\ref{fig:variables_cylinder}),
a region with highly nonlinear, complex flow behavior, is challenging.
Therefore, the physics solver spends a significant
amount of time fixing that region and its speedup is the lowest among all test cases.
Overall, \name\ refines regions
of interest such as near-wall areas (channel flow and flat plate)
and the wake behind the solid body (cylinder),
shows excellent grid convergence properties, and significantly
accelerates the traditional AMR solver by $3.2-5.5\times$. 

\newcommand*\TableVSphysicssolver{
    \begin{table}[htbp]
	\centering
\small
\setlength{\tabcolsep}{7pt}
\renewcommand{\arraystretch}{0.7}
	    \begin{tabular}{l|cc|ccr}

		    & \multicolumn{2}{c|}{\textbf{AMR solver}}
		    & \multicolumn{3}{c}{\textbf{\name}}\\[0.7ex]

		    & ITC
		    & TTC
		    & ITC
		    & \makecell[c]{TTC\\inf + ps}
		    & speedup

		    \\[0.7ex]

\midrule
		    channel flow \Rey\ = \num{2.5e3}
		    & 3369
		    & 3
		    & 2261 
		    & \num{1.2e-2} + 0.68
		    & 4.3$\times$
		    \\[0.7ex]

		    channel flow \Rey\ = \num{15e3}
		    & 4940
		    & 3.1
		    & 2022
		    & \num{1.1e-1} + 0.8
		    & 3.8$\times$
		    \\[0.7ex]

		    flat plate \Rey\ = \num{2.5e5}
		    & 3389
		    & 2.7
		    & 1364
		    & \num{8e-3} + 0.48
		    & 5.5$\times$\\[0.7ex]

		    flat plate \Rey\ = \num{1.35e6}
		    & 5000
		    & 2
		    & 2214
		    & \num{9e-3} + 0.42
		    & 4.7$\times$\\[0.7ex]

		    cylinder \Rey\ = \num{1e5}
		    & 11155
		    & 4.8
		    & 4598
		    & \num{6.3e-3} + 1.5
		    & 3.2$\times$\\[0.7ex]

		    N0012 \Rey\ = \num{2.5e4}
		    & 2267
		    & 2
		    & 1150
		    & \num{5e-3} + 0.55
		    & 3.5$\times$\\[0.7ex]

		    N1412 \Rey\ = \num{2.5e4}
		    & 2637
		    & 2.1
		    & 1720
		    & \num{5e-3} + 0.62
		    & 3.3$\times$\\[0.7ex]
		    \bottomrule
  \end{tabular}
	    \caption{\small Comparison of the time-to-convergence in minutes (TTC) and iterations-to-convergence (ITC) of \name\ and the AMR solver.
	    For \name, we report separately the time spent in inference (inf) and the time spent by the physics solver (ps) driving the solution from inference to convergence, together with the speedup
	    over the AMR solver.\label{tab:vsphysicssolver}}
\end{table}
}

\TableVSphysicssolver

Next, we evaluate \name's performance by comparing it with a baseline neural network. 
Recall that one of the goals of this paper is to perform non-uniform super-resolution
to avoid high-resolution inference in areas that do not require it. 
We hypothesize that \name\ is advantageous over SOTA methods 
that perform uniform super-resolution~\cite{meshfreeflownet,denoisingsuperresolution,surfnet,JFMsuperresolution2021}
because these methods require $64\times$ larger labels for $64\times$ super-resolutions. 
To prove our hypothesis, we build the \baselineNN ~\cite{surfnet} framework and use it as our baseline. 
We compare \name\ with \baselineNN\ using two metrics. 
For a $64\times$ super-resolution, we report, first, the time to achieve the same
accuracy. The \baselineNN's framework also consists of a DNN inference followed by a physics solver that guarantees convergence requirements. 
Hence, we compare both end-to-end frameworks and report
both the inference time and the physics solver time. 
Second, we compare the memory consumption at inference. 
Because both \name\ and \baselineNN\ perform inference on a CPU, we report these metrics on the CPU described in~\ref{sec:experiment:subsec:solvers}. 
The results are presented in Table~\ref{tab:vsbaselinenetwork}.
\newcommand*\TableVSbaselinenetwork{
    \begin{table}[htbp]
	\centering
\small
\setlength{\tabcolsep}{2pt}
\renewcommand{\arraystretch}{0.7}
	    \begin{tabular}{l|ccr|lcr}

		    & \multicolumn{3}{c|}{\textbf{Memory usage}}
		    & \multicolumn{3}{c}{\makecell[c]{\textbf{Time}\\ \textbf{inf + ps} }}\\[1.5ex]

		    & \baselineNN
		    & \name
		    & rf
		    & \baselineNN
		    & \name
		    & speedup
		    \\[0.7ex]
\midrule
		    \makecell[l]{cf \Rey\ = \num{2.5e3}}
		    & 3.9
		    & 0.88
		    & 4.4$\times$
		    & 0.25 + 14
		    & \num{1.2e-2} + 0.68
		    & 20.6$\times$
		    \\[0.7ex]

		    \makecell[l]{cf \Rey\ = \num{15e3}}
		    & 3.9
		    & 0.82
		    & 4.8$\times$
		    & 0.25 + 14.5
		    & \num{1.1e-2} + 0.8
		    & 18.2$\times$
		    \\[0.7ex]

		    \makecell[l]{fp \Rey\ = \num{2.5e5}}
		    & 3.9
		    & 0.62
		    & 6.3$\times$
		    & 0.25 + 11
		    & \num{8e-3} + 0.48
		    & 23$\times$
		    \\[0.7ex]

		    \makecell[l]{fp \Rey\ = \num{1.35e6}}
		    & 3.9
		    & 0.68
		    & 5.7$\times$
		    & 0.25 + 12
		    & \num{9e-3} + 0.42
		    & 28.5$\times$
		    \\[0.7ex]

		    \makecell[l]{cyl \Rey\ = \num{1e5}}
		    & 3.9
		    & 0.52
		    & 7.5$\times$
		    & 0.25 + 10.3
		    & \num{6.3e-3} + 1.5
		    & 7$\times$
		    \\[0.7ex]

		    \makecell[l]{N0012 \Rey\ = \num{2.5e4}}
		    & 3.9
		    & 0.54
		    & 7.2$\times$
		    & 0.25 + 8.35
		    & \num{5e-3} + 0.55
		    & 15.5$\times$
		    \\[0.7ex]

		    \makecell[l]{N1412 \Rey\ = \num{2.5e4}}
		    & 3.9
		    & 0.51
		    & 7.65$\times$
		    & 0.25 + 8.6
		    & \num{5e-3} + 0.62
		    & 14.1$\times$
		    \\[0.7ex]

		    \bottomrule
  \end{tabular}
	    \caption{\small Comparison of \name\ with \baselineNN. Left column compares
	    the GB of memory consumed at each test case's inference and shows
	    the reduction factor (rf) achieved by \name. Right column
	    compares, in minutes, the inference time (inf) and the time to convergence by the physics solver (ps)
	    of both approaches and shows \name's speedup over \baselineNN. cf = channel flow,
	    fp = flat plate, cyl = cylinder, N0012 = NACA0012 (symmetric airfoil),
	    and N1412 = NACA1412 (non-symmetric airfoil).
	    \label{tab:vsbaselinenetwork}}
\end{table}
}

\TableVSbaselinenetwork

Table~\ref{tab:vsbaselinenetwork}'s shows that \name\ significantly outperforms \baselineNN\ for a $64\times$ super-resolution for all test cases. Specifically, we observe $7-28.5\times$ speedups over \baselineNN. We observe the same behavior with the memory usage at inference. 
\baselineNN\ requires almost 4 GB whereas \name\ significantly reduces the memory consumption, realizing $4.4-7.5\times$ reduction factors. 
Note that \name's inference time and memory usage is not consistent through the test cases
because the fine/coarse regions change, as opposed to \baselineNN\ that performs uniform super-resolution. 

\section{Conclusions}
This paper presented \name, a deep learning algorithm that predicts adaptive
mesh refinement. It is an end-to-end framework for non-uniform super-resolution
of turbulent flows that predicts high-resolution accuracy
only in specific regions of the domain while keeping
areas with less complex flow features in the low-precision range
for scalability and performance. 
\name\ is trained with low-resolution data from three different
canonical flows and predicts non-uniform flow fields
for flow cases that have boundary conditions/geometries unseen during training.
\name\ shows excellent discerning properties in all test cases, producing
 higher resolution outputs in regions with complex flow features, such as near-wall 
areas or the wake region behind a cylinder, and keeping low-resolution
patches in those areas that have smooth variations, such as the flow freestream.

\name\ reaches the same convergence guarantees as traditional
AMR solvers, shows excellent quantitative agreement with their heuristics, and accelerates them by $3.2-5.5\times$
in all test cases. Due to its ability to super-resolve only regions
of interest, it reduces the end-to-end time and the memory usage
by $7-28.5\times$ and $4.4-7.65\times$, respectively, over state-of-the-art
DL methods that perform uniform super-resolution.
Future work includes experimenting with different combinations of the number of bins, larger domains with larger patch sizes, different binning strategies, and generalization
to a larger class of geometries and boundary conditions.

\section{Acknowledgement}
This work was partly supported by the National Science Foundation (NSF) under the award number 1750549. Any opinions, findings and conclusions expressed in this material are those of the authors and do not necessarily reflect those of NSF. We thank the HPC3 cluster at the University of California, Irvine, for providing the required hardware to conduct this research.

\small
\bibliographystyle{unsrtnat}
\bibliography{./bib/octavi,./bib/distdl,./bib/FasterLearning,./bib/extra,./bib/vishnu,./bib/apoptosis,./bib/agd,./bib/mathstat,./bib/octavi2,./bib/transformers}

\end{document}